\documentclass[12pt,letterpaper]{article}
\usepackage{graphics}
\usepackage{amssymb,epsfig,amsmath,euscript,array}
\usepackage{axodraw}
\usepackage{subfigure}
\usepackage{color}

\makeatletter
\@addtoreset{equation}{section}
\makeatother

\newcounter{multieqs}
\newcommand{\ii}{{\rm i}}

\newcommand{\fssd}[1]{#1\!\!\!\!/}
\newcommand{\MF}{{\sf B}}
\newcommand{\chiS}{{\lambda}}
\newcommand{\ECH}{{Q_h}}

\def\bd{\begin{document}}
\def\ed{\end{document}}
\def\nn{\nonumber}
\def\bea{\begin{eqnarray}}
\def\eea{\end{eqnarray}}
\let\bm=\bibitem
\let\la=\label

\begin{document}

\hfill{DESY 07-088 }\\[-0.9cm]

\hfill{IPPP 07/32 }\\[-0.9cm]

\hfill{ DCPT 07/64 }\\[-0.9cm]

\hfill{UAB-FT-640}\\[-0.9cm]

\vspace{20pt}

\begin{center}

{\huge \bf  Light from the Hidden Sector\\--\\[1ex]{\large Experimental Signatures of Paraphotons}} \\[1.5ex]

\vspace{30pt}

{\bf M. Ahlers$^1$, H. Gies$^2$, J. Jaeckel$^3$, J. Redondo$^4$ and A. Ringwald$^1$}

{\small \em
{}$^1$Deutsches Elektronen Synchrotron DESY,\\
Notkestrasse 85, D-22607  Hamburg, Germany \\[1ex]
{}$^2$Institute for Theoretical Physics, Heidelberg University,\\
Philosophenweg 16, D-69120 Heidelberg, Germany\\[1ex]
{}$^3$Centre for Particle Theory, Durham University,\\
Durham, DH1 3LE, United Kingdom\\[1ex]
{}$^4$Grup de F{\'\i}sica Te{\`o}rica and Institut
de F{\'\i}sica d'Altes
Energies\\Universitat Aut{\`o}noma de Barcelona, 08193 Bellaterra, Barcelona, Spain}

\vspace{40pt}

{\bf Abstract}

\end{center}

Optical precision experiments are a powerful tool to explore
  hidden sectors of a variety of standard-model extensions with
  potentially tiny couplings to photons.  An important example is given
  by extensions involving an extra light U(1) gauge degree of freedom,
  so-called paraphotons, with gauge-kinetic mixing with the normal
  photon. These models naturally give rise to minicharged particles
  which can be searched for with optical experiments.
In this paper, we study the effects of paraphotons in such
experiments. We describe in detail the role of a magnetic field for
  photon-paraphoton oscillations in models with low-mass minicharged
  particles. In particular, we find that the upcoming
light-shining-through-walls experiments are sensitive to paraphotons
and can distinguish them from axion-like particles.

\setcounter{page}{0}
\thispagestyle{empty}
\newpage

\section{Introduction}

Extensions beyond the current standard model of particle physics often
involve a hidden sector, {\it i.e.}, an experimentally so far unobserved set
of degrees of freedom very weakly coupled to standard-model particles.
Whereas present and future accelerator experiments are generally
devoted to the search for new heavy particles, the potential discovery
of a new weakly coupled light particle requires high-precision
experiments for which non-accelerator setups often appear most
promising.

Optical experiments can provide for such a powerful laboratory tool,
since optical photons can be manipulated and detected with a great
precision. If a hypothetical hidden sector couples effectively 
to photons, optical experiments can have a significant
discovery potential or, alternatively, can put stringent laboratory
bounds on standard-model extensions, since both photon sources and
detectors can be under full experimental control.

An example of such experiments are laser polarization experiments,
such as\linebreak BFRT~\cite{Cameron:1993mr},
PVLAS~\cite{Zavattini:2005tm}, and Q\&A~\cite{Chen:2006cd}, where
linearly polarized laser light is sent through a transverse magnetic
field, and changes in the polarization state are searched for.  The
real and virtual production of
axion-like~\cite{Maiani:1986md,Raffelt:1987im} (ALP) or
minicharged~\cite{Gies:2006ca} (MCP) particles would lead to
observable signals such as an apparent rotation and an ellipticity of
the outgoing laser beam. Similar planned experiments in this direction
are based also on high-intensity lasers \cite{Heinzl:2006xc}.

Another powerful tool are so called light-shining-through-walls (LSW)
experiments, such as BFRT~\cite{Cameron:1993mr,Ruoso:1992nx}.  Here,
laser light is shone onto a wall, and one searches for photons that
appear behind the wall. Vacuum oscillations of photons into
paraphotons with sub-eV masses would lead to a non-vanishing rate of
photons behind the wall~\cite{Okun:1982xi}. In the presence of a
magnetic field, photons can oscillate into axion-like particles, which
can then be reconverted into photons behind the wall by another
magnetic
field~\cite{Sikivie:1983ip,Anselm:1986gz,Gasperini:1987da,VanBibber:1987rq}.\footnote{Also,
  astrophysical observations of light rays from binary pulsar systems
  \cite{Dupays:2005xs}, dimming features in the spectra of TeV gamma
  sources~\cite{Mirizzi:2007hr}, or regenerated photons from cosmic
  ALPs originating from the Crab pulsar~\cite{Fairbairn:2007vj} could
  be a useful optical probe.}

Presently, there are several second generation LSW experiments
worldwide, such as ALPS~\cite{Ehret:2007cm}, BMV~\cite{Rizzo:Patras},
GammeV~\cite{GammeV}, LIPSS~\cite{Baker:Patras},  OSQAR~\cite{OSQAR},
and PVLAS~\cite{Cantatore:Patras}, under construction or serious
consideration (for a review, see
Refs.~\cite{Ringwald:2006rf,Battesti:2007um}).  These efforts are
partially motivated by the report from the PVLAS collaboration of
evidence for a non-zero apparent rotation of the polarization plane of
a laser beam after passage through a magnetic
field~\cite{Zavattini:2005tm}. While the size of the observed effect
greatly exceeds the expectations from quantum
electrodynamics~\cite{Baier,Adler:1971wn,Adler:2006zs,Biswas:2006cr}, it is
compatible with
a photon-ALP oscillation hypothesis or with the production of light
minicharged particles~\cite{Ahlers:2006iz}.  Although the
  couplings and masses required for an explanation of PVLAS seem to be
  in serious conflict with bounds coming from astrophysical
  considerations~\cite{Raffelt:1996,Davidson:2000hf}, there are
  various ways to evade 
  them~\cite{Masso:2005ym,Jain:2005nh,Masso:2006gc,Abel:2006qt,Jaeckel:2006xm,Mohapatra:2006pv,Jain:2006ki,Foot:2007cq,Brax:2007ak,Kim:2007wj} 
  (see, however, Ref.~\cite{Melchiorri:2007sq}).  This makes it
  extremely important to check these interpretations in laboratory
  experiments. 

LSW experiments seem well-suited to distinguish between ALPs and minicharged particles.
Only in the former case, we expect a sizeable rate of regenerated photons. However,
natural models with minicharged particles also contain at least one paraphoton~\cite{Holdom:1985ag}.
In this paper, we include the effects of paraphotons and discuss in detail how this can lead
to a non-vanishing signal in LSW experiments that is nevertheless distinguishable from the one
expected in the ALP case. Moreover, we show that the presence of the paraphoton
significantly alters the signals in polarization experiments.

The paper is organized as follows. In Sect.~\ref{millicharge}, we
briefly review how minicharges arise in models with paraphotons.  In
Sects.~\ref{lsw0} and \ref{lsw1}, we show how paraphotons can lead to
a signal in LSW experiments. In Sect.~\ref{optical}, we discuss how
the predictions for rotation and ellipticity measurements change in
models with paraphotons. In realistic experiments, the magnetic field
region has a finite spatial size.  For small, but non-vanishing
paraphoton mass, this can have significant effects, as we explain in
Sect.~\ref{finite}. In Sect.~\ref{sect:comp}, we give explicit
examples in which signals of a paraphoton model are compared to those
of a pure minicharged particle model without paraphoton.  Furthermore,
we use data from the BFRT experiment to illustrate the sensitivity of
such optical setups. Finally, we summarize and conclude in
Sect.~\ref{conclusions}.

\section{Paraphotons and minicharged particles}\label{millicharge}
Minicharged particles arise very naturally in models with extra U(1)
gauge degrees of freedom
\cite{Okun:1982xi,Holdom:1985ag}. In this section, we
briefly review how kinetic mixing leads to minicharged particles and
provide some details on models that have been proposed to explain the
PVLAS result.

Let us begin with the simplest model with two U(1) gauge groups, one
being our electromagnetic {U(1)$_{_\mathrm{QED}}$}, the other a hidden-sector {U(1)$_\mathrm{h}$} under
which all standard model particles have zero charge. The most general
Lagrangian allowed by the symmetries is
\begin{equation}
{\mathcal{L}}=-\frac{1}{4} F^{\mu\nu}F_{\mu\nu}-\frac{1}{4}B^{\mu\nu}B_{\mu\nu}
-\frac{1}{2}\chi\,F^{\mu\nu}B_{\mu\nu},
\end{equation}
where $F_{\mu\nu}$ is the field strength tensor for the ordinary
electromagnetic {U(1)$_{_\mathrm{QED}}$} gauge field $A^{\mu}$, and
$B^{\mu\nu}$ is the field strength for the hidden-sector
{U(1)$_\mathrm{h}$} field $B^{\mu}$, {\it i.e.}, the paraphoton.  The first
two terms are the standard kinetic terms for the photon and paraphoton
fields, respectively. Because the field strength itself is gauge
invariant for U(1) gauge fields, the third term is also allowed by
gauge and Lorentz symmetry.  This term corresponds to a non-diagonal
kinetic term, a so-called kinetic mixing.

From the viewpoint of a low-energy effective Lagrangian, $\chi$ is a
completely arbitrary parameter. Embedding this into a more fundamental
theory, it is plausible that $\chi=0$ holds at a high-energy scale
related to the fundamental theory. However, integrating out the
quantum fluctuations below this scale generally tends to generate
non-vanishing~$\chi$~\cite{Holdom:1985ag}. In a similar manner, kinetic mixing arises in many
string theory 
models~\cite{Abel:2006qt,Dienes:1996zr,Lust:2003ky,Abel:2003ue,Abel:2004rp,Batell:2005wa,Blumenhagen:2006ux}.

The kinetic term can be diagonalized by a shift
\begin{equation}
\label{shift}
B^{\mu}\rightarrow \tilde{B}^{\mu}-\chi A^{\mu}.
\end{equation}
Apart from a multiplicative renormalization of the gauge coupling,
$e^2\rightarrow e^2/(1-\chi^2)$, the visible-sector fields remain
unaffected by this shift.

Let us now assume that we have a hidden-sector fermion\footnote{Here
  and in the following, we will specialize to the case where the
  hidden-sector particle is a fermion. A generalization to scalars is
  straightforward and does not change the results qualitatively.}  $h$
that has charge one under $B^{\mu}$. Applying the shift~\eqref{shift}
to the coupling term, we find:
\begin{equation}
e_h\bar{h}\fssd{B}\, h\rightarrow e_h\bar{h}\fssd{\tilde{B}}\, h-\chi e_h\bar{h}\fssd{A}\,
 h,
\end{equation}
where $e_h$ is the hidden-sector gauge coupling.
We can read off that the hidden-sector particle now has a charge
\begin{equation}
\label{epsiloncharge}
\epsilon e=-\chi e_h
\end{equation}
under the visible electromagnetic gauge field $A^{\mu}$ which has
gauge coupling $e$. Since $\chi$ is an arbitrary number, the
fractional electric charge $\epsilon$ of the hidden-sector
fermion $h$ is not necessarily integer.

For small $\chi\ll1$, we observe that
\begin{equation}
|\epsilon|\ll 1,
\end{equation}
and $h$ becomes a minicharged particle. From now on we will
concentrate on this case\footnote{Very small values of $\chi$ can be obtained in supersymmetric or string 
theories \cite{Dienes:1996zr}.
On the other hand, light particles with charge  $\epsilon={\mathcal{O}}(1)$ are excluded by several kinds of 
experiments~\cite{Davidson:1991si,Davidson:2000hf} and very
massive particles give negligible contributions in experiments such as
BFRT, PVLAS, Q\&A or the upcoming optical experiments that will test the PVLAS particle interpretation.}.

So far we have considered the case of an unbroken {U(1)$_\mathrm{h}$} symmetry for
the paraphoton. Let us now see what happens if we add a mass
term,\footnote{Adding a mass term is equivalent to breaking the
paraphoton {U(1)$_\mathrm{h}$} via a Higgs mechanism and choosing unitary gauge.}
\begin{equation}
{\mathcal{L}}_\mu=\frac{1}{2}\mu^2 B^{\mu}B_{\mu}.
\end{equation}
Applying the shift \eqref{shift} results in a term
\begin{equation}
\label{masssimple}
{\mathcal{L}}_\mu=\frac{1}{2}\mu^2 \left(\tilde{B}^{\mu}\tilde{B}_{\mu}-2\chi \tilde{B}^{\mu}A_{\mu}+
\chi^2 A^{\mu}A_{\mu}\right)
\end{equation}
that mixes photons with paraphotons.

\begin{figure}[t]
\begin{center}
\scalebox{0.85}[0.85]{
\begin{picture}(190,180)(30,0)
\Photon(0,20)(00,100){5}{7.5}
\Text(-35,60)[l]{\scalebox{1.5}[1.5]{$\gamma$}}
\Text(-45,152)[c]{\scalebox{1.4}[1.4]{$\bar{h}$}}
\Text(45,152)[c]{\scalebox{1.4}[1.4]{$h$}}
\Text(105,152)[c]{\scalebox{1.4}[1.4]{$\bar{h}$}}
\Text(195,152)[c]{\scalebox{1.4}[1.4]{$h$}}
\Vertex(0,100){3}
\Text(15,100)[l]{\scalebox{1.5}[1.5]{$\epsilon e$}}
\ArrowLine(-45,135)(0,100)
\ArrowLine(0,100)(45,135)
\Text(75,75)[c]{\scalebox{2.0}[2.0]{$+$}}
\Photon(150,20)(150,100){5}{7.5}
\Text(165,65)[l]{\scalebox{1.5}[1.5]{$\chi\mu^2$}}
\Text(165,100)[l]{\scalebox{1.5}[1.5]{$e_h$}}
\Text(115,40)[l]{\scalebox{1.5}[1.5]{$\gamma$}}
\Text(115,80)[l]{\scalebox{1.5}[1.5]{$\gamma'$}}
\Vertex(150,100){3}
\ArrowLine(105,135)(150,100)
\ArrowLine(150,100)(195,135)
\Text(230,75)[c]{\scalebox{2.0}[2.0]{$=$}}
\Text(230,90)[c]{\scalebox{0.8}[0.8]{$\gamma$ on shell}}
\Text(260,75)[l]{\scalebox{2.0}[2.0]{$0$}}
\SetWidth{1.5}
\Line(140,70)(160,50)
\Line(160,70)(140,50)
\end{picture}
}
\end{center}
\vspace{-1.0cm} \caption{\small Two diagrams contributing to the
  coupling of the photon to the hidden-sector fermion $h$ in a
  situation where the paraphoton is massive.  The first is the direct
  contribution via the charge $\epsilon e$ that arises from the shift
  \eqref{shift} of the paraphoton field. The second is due to the
  non-diagonal mass term \eqref{masssimple} and cancels the first
  diagram if the external photon is on shell and massless ($q^2=0$).
  Note that the second diagram is only present if the paraphoton has
  non-vanishing mass $\mu^2\neq 0$.}
\label{interaction}
\end{figure}
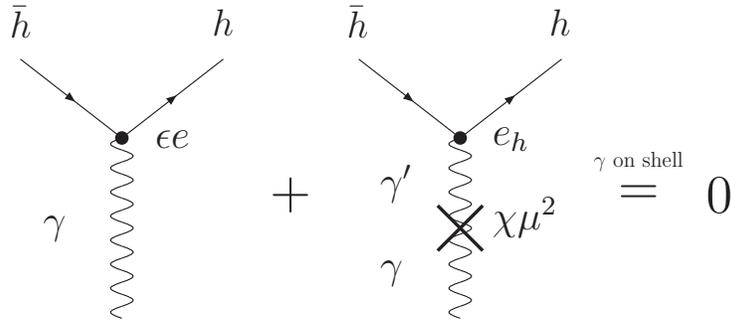

To see how this affects the coupling of the hidden-sector fermion, let
us write down the inverse propagator in our
$(A^{\mu},\tilde{B}^{\mu})$ basis,
\begin{equation}
P^{-1}=\left(
               \begin{array}{cc}
                 q^2-\chi^2\mu^2 & +\chi\mu^2 \\
                 +\chi\mu^2 & q^2-\mu^2 \\
               \end{array}
             \right).
\end{equation}
The effective {charge of the hidden-sector fermion $h$
is now obtained (to lowest order in $\chi$) from}
\begin{equation}
\label{zero}
{\ECH e = \lim_{q^2\to 0}q^2 P_{1j}C_j=-\epsilon e+\chi e_h=0,}
\end{equation}
where
\begin{equation}
C=(-\epsilon e,e_h)
\end{equation}
is the charge vector of $h$ in the $(A^{\mu},\tilde{B}^{\mu})$ basis.
{In this limit, the photon is put} onto the mass shell, and the factor
$q^2$ is {included} to cancel the trivial $1/q^2$ dependence of the
Coulomb potential.  The two contributions {correspond} to the two
diagrams in Fig.~\ref{interaction}.  On shell, the minicharge is
``undone'' by the mass term.  However, off shell or for massive
photons (as, for instance, in a plasma), this is not the case.

Let us now move on to the slightly more involved case of the model
presented in Ref.~\cite{Masso:2006gc} {(``MR model'')}.  This model
involves two paraphotons $B^{\mu}_{1}$ and $B^{\mu}_{2}$. For clarity,
we will in the following suppress Lorentz indices and use a matrix
notation $(A,B_{1},B_{2})$, and similarly for the $\tilde{B}$.  The
Lagrangian for the gauge fields reads:
\begin{equation}
{\mathcal{L}}=-\frac{1}{4}F^{T}{\mathcal{K}}F+\frac{1}{2}A^{T}{\mathcal{M}}A,
\end{equation}
with the kinetic mixing and mass matrix
\begin{equation}
\label{mrmatrices}
{\mathcal{K}}=\left(
  \begin{array}{ccc}
    1 & \chi  & \chi \\
    \chi & 0 & 0 \\
    \chi & 0 & 0 \\
  \end{array}
\right),\quad\quad
{\mathcal{M}}=\left(
\begin{array}{ccc}
        0 & 0 & 0 \\
        0 & \mu^2 & 0 \\
        0 & 0 & 0 \\
\end{array}
\right).
\end{equation}
Again, we can diagonalize the kinetic term by {shifting the
fields},
\begin{eqnarray}
\label{shiftmr}
B_{1}\!\!&\rightarrow&\!\!\tilde{B}_{1}-\chi A,
\\\nonumber
B_{2}\!\!&\rightarrow&\!\!\tilde{B}_{2}-\chi A.
\end{eqnarray}
{This leaves} the ordinary electromagnetic gauge field
unaffected (again up to a small renormalization).

The model of Ref.~\cite{Masso:2006gc} has a hidden-sector fermion that
lives in the bifundamental representation of the two paraphotons with
charges $(0,1,-1)$. Moreover, the two hidden gauge couplings are
assumed to be {equal $e_{{h},1}=e_{{h},2}\equiv e_h$.}
Applying \eqref{shiftmr}, we find
\begin{equation}
\label{cancel}
e_h\bar{h}[B^{\mu}_{1}-B^{\mu}_{2}]\gamma_{\mu}h\rightarrow
e_h\bar{h}[(\tilde{B}^{\mu}_{1}-\chi A^{\mu})-(\tilde{B}^{\mu}_{2}-\chi A^{\mu})]\gamma_{\mu}h
=e_h\bar{h}[\tilde{B}^{\mu}_{1}-\tilde{B}^{\mu}_{2}]\gamma_{\mu}h.
\end{equation}
For the moment, it seems as if the hidden-sector fermion has no
interaction with the photon. However, we should not forget that one of
the paraphotons is massive.  In the new basis, the mass matrix reads
\begin{equation}
\label{massmatrix2}
\tilde{{\mathcal{M}}}=
\left(
  \begin{array}{ccc}
    \chi^2\mu^2 & -\chi\mu^2 & 0 \\
    -\chi\mu^2 & \mu^2 & 0 \\
    0 & 0 & 0 \\
  \end{array}
\right).
\end{equation}
As in the case of only one paraphoton, the mass term undoes the
effects on the minicharges {induced by} the massive paraphoton
(cf.~Eq.~\eqref{zero}).  Since the \emph{second} paraphoton is massless, its
contribution to the minicharge (cf.~middle part of
Eq.~\eqref{cancel}) remains unaffected and the particle has an
effective {charge},
\begin{equation}
\label{effectivecoupling}
 Q_h^{\rm MR}
e=+\chi e_h.
\end{equation}

Finally, let us comment on situations where the virtuality of a
process is high {compared with the paraphoton mass scale}, as,
for instance, in the center of the sun. In this case, we cannot take
the limit $q^2\to 0$ in Eq.~\eqref{zero}. Instead, we have to insert
the virtuality of the process, { implying that} the minicharge
is not undone by the mass term.  At high virtuality, the small mass has
basically no effect and the (first) paraphoton behaves more or less
as if it were massless.  For our case with two paraphotons, this
means that the first paraphoton now contributes a charge
$-e_h\chi$ to the effective electromagnetic coupling of $h$
resulting in a total of
\begin{equation}
 Q_h^{\rm MR} 
\approx 0, \quad\quad  {\rm for}\,\,q^2\gg\mu^2.
\end{equation}
In other words, the mass matrix \eqref{massmatrix2} can be neglected,
and we effectively have the case of two massless paraphotons and an
interaction as in Eq.~\eqref{cancel}.

\section{Light shining through walls I: { \MF =0 }}\label{lsw0}

\begin{figure}[t]
\begin{center}
\begin{picture}(190,80)(0,0)
\Photon(0,60)(100,60){5}{9}
\Photon(100,60)(200,60){5}{9}
\linethickness{0.5cm}
\put(100,20){\line(0,1){80}}
\SetWidth{1.5}
\Line(40,50)(60,70)
\Line(40,70)(60,50)
\Line(140,50)(160,70)
\Line(140,70)(160,50)
\Text(20,75)[c]{\scalebox{1.275}[1.275]{$\gamma$}}
\Text(75,75)[c]{\scalebox{1.275}[1.275]{$\gamma'$}}
\Text(125,75)[c]{\scalebox{1.275}[1.275]{$\gamma'$}}
\Text(180,75)[c]{\scalebox{1.275}[1.275]{$\gamma$}}
\end{picture}
\end{center}
\vspace{-1.0cm} \caption{\small Schematic picture of a
``Light-shining-through-walls'' experiment in absence of a magnetic
field.  The crosses denote the non-diagonal mass terms that convert
photons into paraphotons. The photon $\gamma$ oscillates into the
paraphoton $\gamma'$ and, after the wall, back into the photon
$\gamma$ which can then be detected.}
\label{lswnomagnetic}
\end{figure}
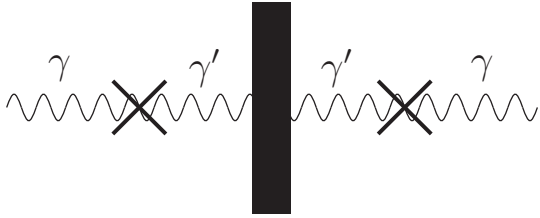

In the previous section, we have seen how a non-diagonal mass matrix
contributes to the effective {charge} of the hidden-sector fermion via
a diagram that changes the photon into a paraphoton {(the second
  diagram of Fig.~\ref{interaction})}.  This non-trivial propagation
of the photon can have interesting effects in itself.  Since the
paraphotons {$\tilde{B}$} do not interact with ordinary matter they
can easily pass through a  wall~\cite{Okun:1982xi},
{giving rise to a process sketched} in Fig.
\ref{lswnomagnetic}.  There, we see how a photon is converted into a
paraphoton by the non-diagonal mass term. Subsequently, the paraphoton
passes through the wall and is then reconverted into an ordinary
photon that can be detected.

{The photon conversion into (massive) paraphotons and back into
  photons very much resembles neutrino oscillations}. Similarly to
  neutrinos, the interaction eigenstates are not equal to the
  propagation eigenstates.

In order to calculate the probability for {an initial photon
  interaction eigenstate to propagate} through a wall via this
  process, we start with the equations of motion in our tilde basis,
\begin{equation}
\label{eom}
[\omega^2\mathbf{1}+\partial^{2}_{z}\mathbf{1}-\tilde{{\mathcal{M}}}]
\left(
                             \begin{array}{c}
                               A \\
                               \tilde{B} \\
                             \end{array}
                           \right)=
\left[(\omega^2+\partial^{2}_{z})\left(
          \begin{array}{cc}
            1 & 0 \\
            0 & 1 \\
          \end{array}
        \right)-\mu^2\left(
               \begin{array}{cc}
                 \chi^2 & -\chi \\
                 -\chi & 1 \\
               \end{array}
             \right)\right]\left(
                             \begin{array}{c}
                               A \\
                               \tilde{B} \\
                             \end{array}
                           \right)=0.
\end{equation}
Here, we have suppressed the Lorentz structure. Both transverse
polarization directions have to fulfill the same equation.  In the
second part, we have specialized to the case of only one massive
paraphoton. {Note that this case} is completely equivalent to the case
with two paraphotons in the model of Ref.~\cite{Masso:2006gc}, because the
mass matrix \eqref{massmatrix2} is non-vanishing only in the first two
components  {and therefore} the second paraphoton does not mix with the photon {(we will see in next section 
that this situation changes when photons propagate in an external magnetic field)}.

\begin{table}[t]\centering
\renewcommand{\arraystretch}{2.0}\small
\begin{tabular}{|l|c|c|c|}
\hline\small
Experiment &\small Laser &\small Cavity &\small Magnets \\
\hline
{\bf\small ALPS}&$532$~nm; $200$~Watt   &$-$&\begin{minipage}[c]{4.cm}\centering $\MF_1=\MF_2=5$~T\\
$\ell_1=\ell_2=4.21$~m\end{minipage}  \\[0.1cm]
\hline
{\bf\small BFRT}& $\sim500$~nm; $3$~Watt&$N_\text{pass}=200$&
\begin{minipage}[c]{4.cm}\centering $\MF_1=\MF_2=3.7$~T\\ $\ell_1=\ell_2=4.4$~m\end{minipage}  \\[0.1cm]
\hline
{\bf\small BMV}& $8\times10^{21}$ $\gamma$ per pulse &$-$&
\begin{minipage}[c]{4.cm}\centering $\MF_1=\MF_2=11$~T\\ $\ell_1=\ell_2=0.25$~m\end{minipage}  \\[0.1cm]
\hline
{\bf\small GammeV}&$532$~nm; $3.2$~Watt    &$-$&
\begin{minipage}[c]{4.cm}\centering $\MF_1=\MF_2=5$~T\\ $\ell_1=\ell_2=3$~m\end{minipage}  \\[0.1cm]
\hline
{\bf\small LIPSS}&$900$~nm; $3000$~Watt    &$-$&
\begin{minipage}[c]{4.cm}\centering $\MF_1=\MF_2=1.7$~T\\ $\ell_1=\ell_2=1$~m\end{minipage}  \\[0.1cm]
\hline
{\bf\small OSQAR}&$1064$~nm; $1000$~Watt  &$N_\text{pass}\sim10000$&
\begin{minipage}[c]{4.cm}\centering$\MF_1=\MF_2=9.5$~T\\$\ell_1=\ell_2=7$~m\end{minipage} \\[0.1cm]
\hline
{\bf\small PVLAS}& $1064$~nm; $0.02$~Watt&$N_\text{pass}=44000$&
\begin{minipage}[c]{4.cm}\centering$\MF_1=5$~T, $\ell_1=1$~m\\ $\MF_2=2.2$~T, 
$\ell_2=0.5$~m\end{minipage}\\[0.1cm]
\hline
\end{tabular}
\caption{\small The benchmark values of ``light-shining-through-walls''
  (LSW) experiments (for some of these experiments, the setup is still preliminary).}\label{tab1}
\end{table}

\begin{figure}[t]
\begin{center}
\includegraphics[width=0.7\linewidth]{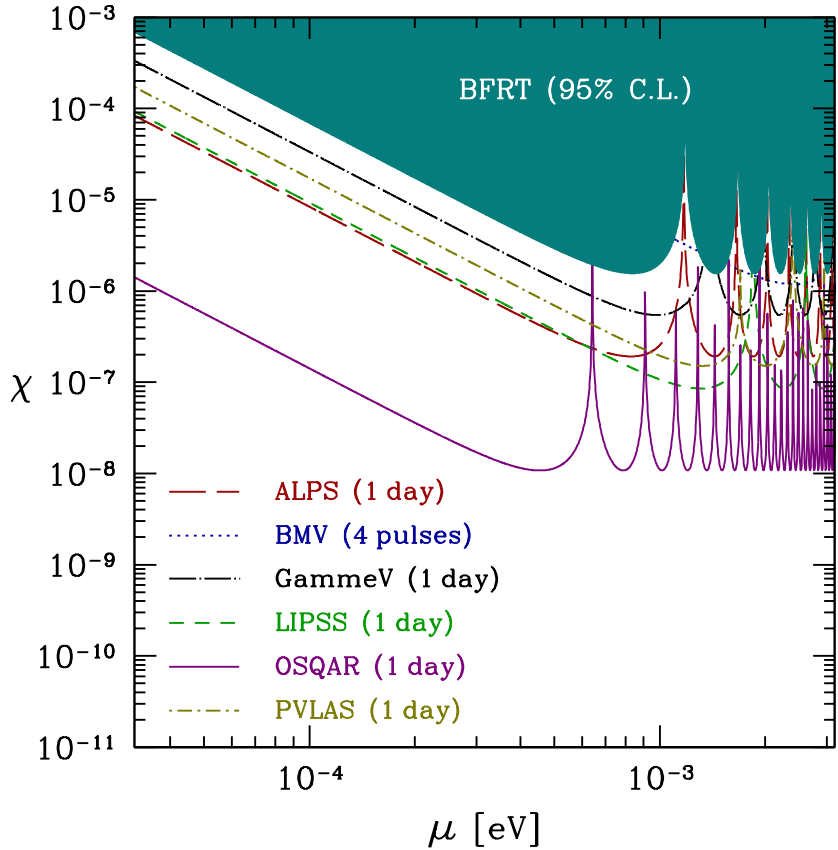}
\end{center}
\caption{\small Projected sensitivity (one expected event per indicated time; no background; 
 $\eta =1$) 
of future LSW
experiments {to photon-paraphoton oscillations  in the absence of 
  a magnetic field.}  The shaded region shows the $95\%$ exclusion
region of BFRT.}\label{fig1}
\end{figure}

From Eq.~\eqref{eom}, we find two propagating eigenstates,
\begin{eqnarray}
\label{propeigenstate}
V_{1}(z,t)\!\!&=&\!\!\left(
                              \begin{array}{c}
                                1 \\
                                \chi \\
                              \end{array}
                            \right)
\exp(-\ii(\omega t\pm k_{1}z)),\quad\,\,\, {\rm{with}} \quad k^2_{1}=\omega^2,
\\\nonumber
V_{2}(z,t)\!\!&=&\!\!\left(
                  \begin{array}{c}
                    -\chi \\
                    1 \\
                  \end{array}
                \right)
\exp(-\ii(\omega t\pm k_{2}z)),\quad {\rm{with}} \quad k^2_{2}=\omega^2-\mu^2+{\mathcal{O}}(\chi^2).
\end{eqnarray}
Let us now start with an initial state at $z=0$ that is purely
photonic\footnote{In Appendix \ref{initialstate}, we argue that
    this is a reasonable choice for the initial state.},
\begin{equation}
A(0,0)=A_{0}\left(
  \begin{array}{c}
    1 \\
    0 \\
  \end{array}
\right)=
A_{0}\left(\frac{1}{1+\chi^2}V_{1}(0,0)-\frac{\chi}{1+\chi^2}V_{2}(0,0)\right).
\end{equation}
The survival probability for an initial photon is
\begin{equation}
\label{survival}
P_{\gamma\to\gamma}(z)=\frac{|A_{1}(z,t)|^2}{|A_{0}|^2}=1-4\chi^2\sin^{2}\left(\frac{{\Delta k} \,
  z}{2}\right)+{\mathcal{O}}(\chi^4),
\end{equation}
where
\begin{equation}
\label{delay}
{\Delta k}=k_{1}-k_{2}\approx \frac{\mu^2}{2\omega},\quad{\rm{for}}\quad \mu\ll\omega.
\end{equation}
The conversion probability into paraphotons  is then obtained as~\cite{Okun:1982xi}
\begin{equation}
\label{transition}
P_{\gamma\to\gamma'}(z)=\frac{|A_{2}(z,t)|^2}{|A_{0}|^2}=1-\frac{|A_{1}(z,t)|^2}{|A_{0}|^2}=
4\chi^2\sin^{2}\left(\frac{\mu^2 }{4\omega}z\right).
\end{equation}
In a light-shining-through-walls experiment as depicted in
Fig.~\ref{lswnomagnetic},  with lengths $\ell_{1}$ and $\ell_{2}$
before and after the wall, the photon {probability for a transit}
``through the wall'' is then simply given by
{\begin{equation}\label{P1}
P_\text{trans} =P_{\gamma\to\gamma'}(\ell_{1})P_{\gamma'\to\gamma}(\ell_{2})= 
16 \chi^4\left[\sin\left(\frac{\ell_1 \mu^2}{4\omega}\right)
\sin\left(\frac{\ell_2 \mu^2}{4\omega}\right)\right]^2.
\end{equation}
Typically, the conversion probability
$P_{\gamma\to\gamma'}(\ell_{1})$ is enhanced by using a pair of
facing mirrors before the wall. If the photon beam is reflected
$N_{\rm{pass}}$ times, it will make \linebreak $(N_{\rm{pass}}+1)/2$ ``attempts''
to cross the wall, enhancing the transmission probability by this same
factor.}
The expected rate of observed photons  {in addition
involves} the total  initial photon rate $N_0$ and the detection efficiency
$\eta<1$,
\begin{equation}
N =  \eta N_0\left[ { \frac{N_{\rm{pass}}+1}{2}}\right] P_\text{trans}\,.
\end{equation}
Figure \ref{fig1} shows the limit from the BFRT
experiment~\cite{Cameron:1993mr} and the projected sensitivity of the
on-going experiments listed in Table~\ref{tab1}, corresponding to one
regenerated photon after one day of observation. For $\mu\gtrsim
10^{-4}$~eV, this limit on the mixing parameter is better than the one
from Cavendish-type laboratory searches for a fifth
force~\cite{Williams:1971ms,Bartlett:1988yy,Popov:1999}.

\section{Light shining through walls II: \MF$\neq 0$}\label{lsw1}

In a classic light-shining-through-walls 
experiment~\cite{Anselm:1986gz,Gasperini:1987da,VanBibber:1987rq}, the
light is shone through a {transverse} magnetic field.  
This is because these experiments typically look for axions
\cite{Wilczek:1977pj,Weinberg:1977ma}, whose production by
virtue of their coupling to two photons requires a {transverse}
magnetic field~\cite{Sikivie:1983ip}. Therefore, let us
study what happens in our photon-paraphoton system with a minicharged
particle if we switch on such a magnetic field.

In a pure minicharged particle model without paraphotons, {the
  conversion of photons into minicharged particles in a magnetic
  field} does not {produce a photon signal in the detector behind the
  wall}. The particle-antiparticle pairs that are created from the
photons \cite{Gies:2006ca} move away from each other under the
influence of the magnetic field because they have opposite charges.
Moreover, they typically have opposite momenta {along the} direction
of the magnetic field lines separating them even further. In general,
the particles will not { annihilate} again behind the wall and cannot
be reconverted into photons\footnote{In the present work, we assume
  that the wall thickness is bigger than the Compton wavelength of the
  minicharged particles. In this limit, we expect that the potential
  process of a photon propagating through the wall as a virtual
  minicharged particle pair is exponentially suppressed. The opposite
  limit requires a careful field-theoretical study of the photon
  polarization tensor near the wall, which is beyond the scope of the
  present work.}.

\begin{figure}[t]
\begin{center}
\subfigure[]{
\scalebox{0.85}[0.85]{
\begin{picture}(190,120)(0,40)
\Photon(0,90)(70,90){5}{7.5}
\Text(30,110)[c]{\scalebox{1.5}[1.5]{$\gamma$}}
\Text(160,110)[c]{\scalebox{1.5}[1.5]{$\gamma$}}
\Text(40,73)[l]{\scalebox{1.5}[1.5]{$\epsilon e$}}
\Text(140,73)[l]{\scalebox{1.5}[1.5]{$\epsilon e$}}
\CArc(100,90)(30,0,180)
\CArc(100,90)(30,180,360)
\CArc(100,90)(27,0,180)
\CArc(100,90)(27,180,360)
{\SetWidth{3}
\ArrowLine(100,118)(106,118)}
\SetOffset(0,0)
\Photon(130,90)(200,90){5}{7.5}
\end{picture}
}
\label{convertAA}}
\hspace{1cm}
\subfigure[]{
\scalebox{0.85}[0.85]{
\begin{picture}(190,120)(0,40)
\Photon(0,90)(70,90){5}{7.5}
\Text(30,110)[c]{\scalebox{1.5}[1.5]{$\gamma$}}
\Text(160,110)[c]{\scalebox{1.5}[1.5]{$\gamma'$}}
\Text(40,73)[l]{\scalebox{1.5}[1.5]{$\epsilon e$}}
\Text(140,73)[l]{\scalebox{1.5}[1.5]{$e_\mathrm{h}$}}
\CArc(100,90)(30,0,180)
\CArc(100,90)(30,180,360)
\CArc(100,90)(27,0,180)
\CArc(100,90)(27,180,360)
{\SetWidth{3}
\ArrowLine(100,118)(106,118)}
\SetOffset(0,0)
\Photon(130,90)(200,90){5}{7.5}
\end{picture}
}
\label{convertBB}}
\subfigure[]{\begin{picture}(190,100)(0,20)
\SetScale{1}
\SetOffset(-100,90)
\Line(0,0)(70,00)
\Line(0,2)(70,2)
{\SetWidth{3} \ArrowLine(34,1)(39,1)}
\Text(75,0)[l]{\scalebox{1.5}[1.5]{$=$}}
\SetOffset(00,90)
 \ArrowLine(0,0)(60,0)  \Text(65,0)[l]{\scalebox{1.5}[1.5]{$+$}}
\SetOffset(85,90)
 \ArrowLine(0,0)(30,0) \ArrowLine(30,0)(60,0) \Text(65,0)[l]{\scalebox{1.5}[1.5]{$+$}}
 \Photon(30,0)(30,-40){5}{5} \CArc(30,-47)(7,0,360) \Line(35,-52)(25,-42) \Line(35,-42)(25,-52) 
\Text(38,-55)[l]{\scalebox{1.2}[1.2]{$\mathbf{\MF}$}}
\SetOffset(170,90)
 \ArrowLine(0,0)(30,0) \ArrowLine(30,0)(60,0)  \Photon(30,0)(30,-40){5}{5} \CArc(30,-47)(7,0,360) 
\Line(35,-52)(25,-42)
 \Line(35,-42)(25,-52)  \Text(38,-55)[l]{\scalebox{1.2}[1.2]{$\mathbf{\MF}$}}
\SetOffset(200,90)
\ArrowLine(0,0)(30,0) \ArrowLine(30,0)(60,0)  \Photon(30,0)(30,-40){5}{5} \CArc(30,-47)(7,0,360) 
\Line(35,-52)(25,-42) \Line(35,-42)(25,-52)
\Text(38,-55)[l]{\scalebox{1.2}[1.2]{$\mathbf{\MF}$}}
 \Text(70,0)[l]{\scalebox{1.5}[1.5]{$+\ ...$}}
\SetOffset(0,0)
\Text(92,78)[l]{\scalebox{1.275}[1.275]{{$\epsilon e$}}}
\Text(177,78)[l]{\scalebox{1.275}[1.275]{{$\epsilon e$}}}
\Text(238,78)[l]{\scalebox{1.275}[1.275]{{$\epsilon e$}}}
\end{picture}
\label{convertCC}}
\vspace{.1cm}
\end{center}
\vspace{-0.5cm} \caption{\small {
    {The contribution of minicharged particles to the
      polarization tensor \ref{convertAA}. The real part leads to
      birefringence, whereas the imaginary part reflects the absorption
      of photons caused by the production of particle-antiparticle
      pairs.}  The analogous diagram \ref{convertBB} shows how
    minicharged particles mediate transitions between photons and
    paraphotons. Note that the latter diagram is enhanced with respect
    to the first one by a factor {$\sim e_\mathrm{h}/(\epsilon e){=1/\chi}$}.
    The double line represents the complete propagator of the
    minicharged particle in an external magnetic field $\mathbf{\MF}$
    as displayed in \ref{convertCC} \cite{Schwinger:1951nm}. }}
\label{convert-NP}
\end{figure}
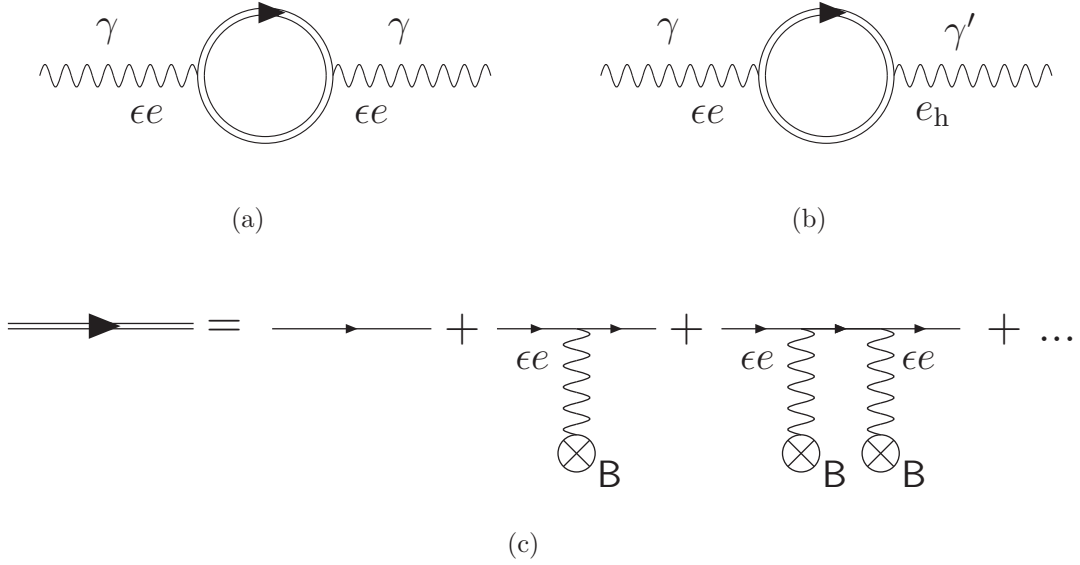

What is different if we include paraphotons?  The {\emph{big}}
difference is that now photons can convert into paraphotons which then
will pass through the wall.  In the presence of a magnetic field, this
coherent conversion {is possible even} for massless paraphotons.
The relevant diagram for this transition is depicted in Fig.
\ref{convertBB}.

For a quantitative analysis, we again start from the equations of
motion. We begin with the simple case of only one massless paraphoton.
Without paraphotons, Fig.~\ref{convertAA} {would induce a
  non-vanishing refractive index. The photon would then satisfy the
  following equation of motion,}
\begin{equation}
\left[(1+2\epsilon^{2} e^2\Delta
  N_{i})\omega^{2}-k^2\right]A_{i}=0 \quad\quad({\rm{without}}\,\,{\rm{paraphotons}}).
\end{equation}
The index $i$ represents the two polarizations $\parallel$ and $\perp$
with respect to the magnetic field and $\epsilon^{2} e^2\Delta
N_{i}=n_{i}-1$ is the contribution to the refractive index of the
photon caused by the diagram \ref{convertAA}.  The explicit expression
for $\Delta N_{i}(\epsilon e\MF,m_{f})$ for a particle $h$ with mass
$m_{\epsilon}$ is given in Appendix \ref{refraction}. {Various
  representations of $\Delta N_i$ and plots of the parameter
  dependencies can, for instance, be found in
  \cite{Erber:1966vv,Dittrich:2000zu,Shore:2007um}. }

It is now straightforward to {derive} the contribution from Fig.
\ref{convertBB} to our photon-paraphoton {system. The full equation of
  motion becomes}
\begin{equation}
\label{photoneom1}
\left[(1+2\epsilon^{2} e^2\Delta N_{i})\omega^{2}-k^2\right]A_{i}+
2(\epsilon e)e_h\Delta N_{i} \,\omega^2\tilde{B}_{i}=0.
\end{equation}

Equation \eqref{photoneom1} is not a closed equation for the photon,
because it contains the paraphoton field. The equation of motion for
the paraphoton can be obtained in complete analogy. We simply have to
replace the two external photon legs in Fig.~\ref{convertAA} with
paraphotons and exchange the photon and the paraphoton field in Fig.
\ref{convertBB},
\begin{equation}
\label{photoneom2}
\left[(1+2{e^{2}_{h}}\Delta N_{i})\omega^{2}-k^2\right]\tilde{B}_{i}+
2(\epsilon e)e_h\Delta N_{i}\,\omega^2A_{i}=0.
\end{equation}
Using \eqref{epsiloncharge} to eliminate $\epsilon$, we can write
the complete set of equations as
\begin{equation}
\left[(\omega^{2}+\partial^{2}_{z})
\left(
  \begin{array}{cc}
    1 & 0 \\
    0 & 1 \\
  \end{array}
\right)
+2\omega^2{e^{2}_{{h}}}\Delta N_{i}\left(
  \begin{array}{cc}
    +\chi^2  & -\chi  \\
    -\chi  & +1 \\
  \end{array}
\right)\right]\left(
                \begin{array}{c}
                  A_{i} \\
                  \tilde{B}_{i} \\
                \end{array}
              \right)=0,
\end{equation}
for one massless paraphoton.

This equation is completely equivalent to \eqref{eom} if we replace
\begin{equation}
\label{replace}
\mu^2\rightarrow -2\omega^{2}{e^{2}_{{h}}}\Delta N_{i}.
\end{equation}
The propagation eigenstates are already given in
Eq.~\eqref{propeigenstate}. There is only one slight complication that
{has to be dealt with} when calculating the transition
probability: $\Delta N_{i}$ is generally complex,
\begin{equation}
\Delta N_{i}=\Delta n_{i}+\frac{1}{2\omega}\ii \kappa_{i}.
\end{equation}
{Up to coupling factors corresponding to the external
  (para-)photon lines,} $\Delta n_{i}$ is the deviation of the real
refractive index from $1$, and $\kappa_{i}$ denotes the absorption
coefficient.  Accounting for this, the transition probability is
\begin{equation}
\label{transitionmagnetic}
P_{\gamma\to\gamma'}(z)=\chi^2[1+\exp(- e^{2}_{{h}}\kappa_{i} z)-
2\exp(- e^{2}_{{h}}\kappa_{i} z/2){\cos(e^{2}_{h}\Delta n_{i} \omega z)}].
\end{equation}
The total probability for a light-shining-through-walls experiment
then is
\begin{equation}
\label{transmagnetic}
P_\text{trans} = \left[
\frac{N_\text{pass}+1}{2}\right]\chi^4[1+\exp(-
  e^{2}_{{h}}\kappa_{i} {z})-2\exp(-
  e^{2}_{{h}}\kappa_{i} z /2){\cos(e^{2}_{h}\Delta n_{i} \omega z)}]^{2}.
\end{equation}

Note the following features:
\begin{itemize}
\item{} The size of the photon-paraphoton mixing is controlled by
$\chi^2$, but

\item{} the typical oscillation length for the photon-paraphoton
system is given by \linebreak
{$1/(\omega {e^{2}_{{h}}}\Delta n)$}. The latter is by a factor $\chi^2$ shorter
than the typical length which might naively be expected from
Fig.~\ref{convertAA}.

\item{} The oscillations die out for non-vanishing $\kappa_{i}$ and we
get a non-oscillatory signal for experiments with a sufficiently long
conversion region.  This is rather useful, because the oscillations
typically lead to  ``holes''  in the sensitivity of the experiment for a
given fixed experimental signal.

\item{} The $\Delta N_{i}$ are non-vanishing for both polarization
  directions $\parallel$ and $\perp$ and we expect a signal for both
  polarizations. {This might resemble a case in which
    light-{shining}-through-walls proceeds through an axion-like
    particle (ALP) with mixed parity interactions to
    photons\footnote{{Even in this case it can be easy to distinguish
        between a general ALP and paraphoton scenarios. In the second
        case, the ratio {of the regeneration rates of the two
          polarization modes} does depend on the
        photon energy and on the strength of the magnetic field,
        whereas this ratio is a constant for the former case.}}, as
    considered in Ref.~\cite{Liao:2007nu}. However, for the most
    likely scenarios where the ALP has a definite parity, either
    pseudoscalar or scalar, a signal would be expected only for the
    $\parallel$ \emph{or} $\perp$ mode.}

\item{} For practical {purposes}, it is useful that the oscillation
  length of the photon-parapho\-ton system is controllable by the
  external magnetic field ($\Delta n$ and $\kappa$ depend on $\MF$).
  Varying the magnetic field, one can try to maximize the term
    in square brackets in Eqs.~\eqref{transitionmagnetic} or
    \eqref{transmagnetic}. For instance, the transition probability
    \eqref{transitionmagnetic} asymptotically approaches $\chi^2$; but
    for a suitable set of parameters such that $\kappa_i z\to 0$ and
    $\Delta n_i \omega z\to \pi$, the transition probability can
    increase up to $16\chi^2$. This is in contrast to the case of an
  ALP, where the oscillation length is completely determined by the
  mass of the ALP, which cannot be changed, {and the frequency of the
  laser,} which is at least more difficult to change.
\end{itemize}

{At first glance, the insertion of a mass term seems
straightforward on the basis of the equations of motion}. However, as
discussed in Sect.~\ref{millicharge}, we have to take into account
that the effective $h$-photon coupling receives an additional
contribution from the non-diagonal propagator, {such that}
$\ECH=0$.  Therefore, {$\Delta N_{i}$} vanishes in
this case, and we get the same result as for $\MF=0$. Note that this
simple argument implicitly assumes that the magnetic field is
homogeneous and thus has infinite spatial extent, also transversally to
the photon beam direction. The effects of a magnetic field with finite size will
be discussed in {Sect.~\ref{finite}.}

Finally, let us turn to the full model \cite{Masso:2006gc} with two
paraphotons, one massless and one massive.
As discussed in Sect.~\ref{millicharge}, the effective coupling of the particle $h$ with charges $(0,1,-1)$
to photons is $\ECH e=\chi e_h$, cf.~Eq.~\eqref{effectivecoupling}.
This determines $\Delta N$ as given in  { Section \ref{finite} and Appendix \ref{refraction}}.
Taking the negative
charge of $h$ with respect to the second paraphoton into account, the
equation of motion reads
\begin{eqnarray}
\label{complicated}
&&\!\!\!\!\!\!\!\!\!\!\!\!\!\!\!\!\!\!\!\Bigg[(\omega^{2}+\partial^{2}_{z})\left(
                                     \begin{array}{ccc}
                                       1 & 0 & 0 \\
                                       0 & 1 & 0 \\
                                       0 & 0 & 1 \\
                                     \end{array}
                                   \right)
-\mu^2\left(
  \begin{array}{ccc}
    \chi^2 & -\chi & 0 \\
    -\chi & 1 & 0 \\
    0 & 0 & 0 \\
  \end{array}
\right)
\\\nonumber
&&\quad\quad\quad\quad\quad\quad\quad\quad\quad\quad\quad\quad\quad+2\omega^2 e^{2}_{{h}} \Delta N_{i}\left(
                                           \begin{array}{ccc}
                                              0   & 0 & 0 \\
                                              0     & 1 & -1 \\
                                              0 & -1 & 1 \\
                                           \end{array}
                                         \right)
\Bigg]\left(
         \begin{array}{c}
           A_{i} \\
           \tilde{B}_{1,i} \\
           \tilde{B}_{2,i} \\
         \end{array}
       \right)
=0.
\end{eqnarray}
The explicit regeneration probabilities are given in Appendix
\ref{MRregen}. A quantitative discussion follows below in Sect.~\ref{sect:comp}.

\section{Dichroism and birefringence in models with paraphotons}\label{optical}

In the preceding sections, we have concentrated on
light-shining-through-walls experiments. But imprints of paraphotons
can also be found in experiments that measure the change in the
optical properties after propagation through the {apparatus,
as is, for instance, done in the BFRT, PVLAS and Q\&A
experiments.}

Both rotation and ellipticity can be {inferred from the
photon-photon} amplitude,
\begin{equation}
\label{amplitude}
A^{i}_{\gamma\to\gamma}=\frac{A^{i}_{1}(z,t)}{A_{0}\exp(\ii (kz-\omega t))},
\end{equation}
for different polarization directions $i$.

As we have already seen in Sect.~\ref{lsw0}, Eq.~\eqref{survival},
\begin{equation}
P^{i}_{\gamma\to\gamma}=|A^{i}_{\gamma\to\gamma}|^{2}
\end{equation}
is the survival probability for an incoming photon. In other words,
{$1-|A^{i}_{\gamma\to\gamma}|$} is the decrease in amplitude for the
different polarization directions.  From this, we can easily find the
rotation of an initially linear polarized beam entering at an angle
$\theta$,
\begin{equation}
\label{rotation}
\Delta \theta=\frac{1}{2}(|A^{\perp}_{\gamma\to\gamma}|-|A^{\parallel}_{\gamma\to\gamma}|)\sin(2\theta)
\approx \frac{1}{2}{\rm{Re}}(A^{\perp}_{\gamma\to\gamma}-A^{\parallel}_{\gamma\to\gamma})\sin(2\theta),
\end{equation}
where the approximation is valid for amplitudes that are close to $1$.

Phase shifts compared to an unmodified photon beam appear as the
argument of the amplitude,
${\rm{Arg}}(A^{\perp,\parallel}_{\gamma\to\gamma})$.  One finds for
the ellipticity,
\begin{equation}
\label{ellipticity}
\psi=\frac{1}{2}[{\rm{Arg}}(A^{\parallel}_{\gamma\to\gamma})-
{\rm{Arg}}(A^{\perp}_{\gamma\to\gamma})]\sin(2\theta)
\approx \frac{1}{2}{\rm{Im}}(A^{\parallel}_{\gamma\to\gamma}-A^{\perp}_{\gamma\to\gamma})\sin(2\theta).
\end{equation}

As expected, {neither rotation nor ellipticity appears in the
absence} of a magnetic field, because the amplitudes
$A^{\parallel,\perp}_{\gamma\to\gamma}$ are equal. This is, of course,
due to the fact that a simple Lorentz invariant mass term {distinguishes} no
preferred direction.

In the presence of a magnetic field, however, the amplitudes differ,
because the oscillation and absorption lengths are different for
photons parallel $\parallel$ and perpendicular $\perp$ to the magnetic
field.

Using the propagation eigenstates derived in Sects. \ref{lsw0} and
\ref{lsw1}, namely Eqs.~\eqref{propeigenstate}, \eqref{delay} and
\eqref{replace}, we find the amplitude
\begin{equation}
A^{\parallel,\perp}_{\gamma\to\gamma}=1-\chi^2 
(1-\exp(-\ii \Delta{k}^{\parallel,\perp} z-K^{\parallel,\perp} z)),\quad {\rm{for}}\quad \chi\ll 1,
\end{equation}
where
\begin{equation}\label{defKD}
\Delta{k}^{\parallel,\perp}=-\omega e^{2}_{{h}}\Delta n^{\parallel,\perp}, 
\quad {K^{\parallel,\perp}=\frac{1}{2} e^{2}_{{h}}\kappa^{\parallel,\perp}}.
\end{equation}

Inserting this into Eq.~\eqref{rotation}, we find:
\begin{eqnarray}\label{rotKD}
\Delta \theta\!\!&=&\!\!\frac{1}{2}\chi^2
\left[ \cos(\Delta{k} ^{\perp} z)\exp(-K^{\perp} z)-\cos(\Delta{k} ^{\parallel} z)
\exp(-K^{\parallel} z)\right]\sin(2\theta)
\\\nonumber
\!\!&\approx&\!\!\left[\frac{1}{4} \epsilon^{2} e^{2} (\kappa^{\parallel}-\kappa^{\perp}) z
+\frac{1}{4} \chi^2 \omega^{2}
[(e_h^2 \Delta n^{\parallel})^{2}-(e_h^2 \Delta n^{\perp})^{2}]z^2
\right]\sin(2\theta),\quad{\rm{for}}\,\, \Delta{k} z,K z\ll 1.
\end{eqnarray}
The first term in the last line is the standard result for the
rotation in a model without paraphotons (cf.~{\it e.g.}~Ref.~\cite{Ahlers:2006iz}). However, note that with paraphotons
where $\epsilon^2 e^2=\chi^2 e_h^2$ this result holds only if
the length $z$ is much smaller than the oscillation length $1/(\omega
e^2_{{h}}\Delta n)$; the latter is by a factor $\chi^2$ smaller than
the naive expectation from the case without paraphotons $1/(\omega
\epsilon^{2} e^{2} \Delta n)$.

Similarly the ellipticity can be inferred from
Eq.~\eqref{ellipticity},
\begin{eqnarray}\label{ellKD}
\psi\!\!&=&\!\!-\frac{1}{2}\chi^2 \left[\sin(\Delta{k} ^{\parallel} z)\exp(-K^{\parallel} z)-
\sin(\Delta{k}^{\perp} z)\exp(-K^{\perp} z)\right]\sin(2\theta)
\\\nonumber
\!\!&\approx&\!\!\frac{1}{2}{\omega} \epsilon^{2}_{h} e^{2} (\Delta n^{\parallel}-\Delta n^{\perp}) 
z\sin(2\theta), \quad\quad{\rm{for}}\quad \Delta{k} z,K z\ll 1.
\end{eqnarray}

Eqs.~\eqref{rotation} and \eqref{ellipticity} are valid also for
models with two paraphotons. The determination of the rotation and
ellipticity boils down to solving the equation of motion
\eqref{complicated} and inserting into \eqref{amplitude},
\eqref{rotation} and \eqref{ellipticity}.
(The necessary expressions for the amplitudes can be found in Appendix \ref{MRregen}.) 
A quantitative discussion follows in {Sect.~\ref{sect:comp}.}

\section{Effects of a magnetic field with finite extent transverse to
  the photon beam}\label{finite}

In Sect.~\ref{millicharge}, we have seen that, for massive paraphotons,
the $\epsilon$ electric charge resulting from the shift in the
paraphoton field is effectively canceled by the mass term as depicted
in Fig.~\ref{interaction}. However, this is true only if the photon
coupling to the hidden-sector particle has $q^2=0$, {\it i.e.}, if it is
on shell. 

In realistic situations, the magnetic background field has a finite
extent and the photons {which build {it} up} have
a non-vanishing virtuality.  In order to take this into account, we
have to resum the diagrams in Fig.~\ref{interaction} also at
non-vanishing virtuality. Resumming tree-level diagrams is equivalent
to solving the equations of motion (this automatically includes also
the higher-order diagrams with multiple mass insertions that were
neglected in Fig.~\ref{interaction}).  Therefore, we have to solve the
combined equations of motion for photon and paraphoton --
{including} the mass term -- not only for the photons of the laser but
also for the background magnetic field. To lowest order, we can
neglect the index of refraction $\Delta N$ and we have (Lorentz
structure suppressed),
\begin{equation}
\label{eom2}
[\nabla^2 \mathbf{1}-\tilde{{\mathcal{M}}}]
\left(
                             \begin{array}{c}
                               A \\
                               \tilde{B} \\
                             \end{array}
                           \right)=
\left[\nabla^2\left(
          \begin{array}{cc}
            1 & 0 \\
            0 & 1 \\
          \end{array}
        \right)-\mu^2\left(
               \begin{array}{cc}
                 \chi^2 & -\chi \\
                 -\chi & 1 \\
               \end{array}
             \right)\right]\left(
                             \begin{array}{c}
                               A \\
                               \tilde{B} \\
                             \end{array}
                           \right)=0,
\end{equation}
for a static background field.

To get an impression of the general behavior, we can solve
\eqref{eom2} for a spherically symmetric situation with a point
source.  Similar to the two eigenmodes in Sect.~\ref{lsw0}, we find
two solutions corresponding to a pure massless Coulomb-type potential
and a massive Yukawa-type potential\footnote{In our simplified
notation without any Lorentz structure, the potentials can be either
the electric potential or the vector potential leading to magnetic
fields, depending on whether the source is a charge or a current.},
\begin{eqnarray}
\label{propeigenstate2}
\phi_{1}(r)\!\!&=&\!\!\left(
                              \begin{array}{c}
                                1 \\
                                \chi \\
                              \end{array}
                            \right)
\frac{1}{r},
\\\nonumber
\phi_{2}(r)\!\!&=&\!\!\left(
                  \begin{array}{c}
                    -\chi \\
                    1 \\
                  \end{array}
                \right)
\frac{\exp\left(-\mu r\right)}{r}.
\end{eqnarray}
For a source made up of ordinary matter, the potentials have to behave
like $\sim (1,0)^{T} 1/r$ for $r\to 0$ and the potential for matter
fields takes the form,
\begin{equation}
\phi_{\rm{matter}}\sim \frac{1}{1+\chi^2}\frac{1}{r}\left(
                                                      \begin{array}{c}
                                                        1+\chi^2\exp(-\mu r) \\
                                                        \chi(1-\exp(-\mu r)) \\
                                                      \end{array}
                                                    \right).
\end{equation}
A hidden-sector particle with charge vector $(\epsilon e,
e_h)^{T}$ therefore sees an effective potential,
\begin{equation}
(\epsilon e, e_h)\phi_{\rm{matter}}\sim 
\frac{1}{r} [\epsilon e+\chi e_h(1-\exp(-\mu r))]+{\mathcal{O}}(\chi ^2)
=\frac{1}{r}\epsilon e \exp(-\mu r),
\end{equation}
where we have used Eq.~\eqref{epsiloncharge}, $\epsilon e=-e_h\chi$, for the last equality.  
Note that this can be
  written as
\begin{equation}
(\epsilon e, e_h)\phi_{\rm{matter}} \sim 
\left.(\epsilon e, e_h)\phi_{\rm{matter}}\right|_{\mu=0}  \exp(-\mu r),
\end{equation}
and therefore, these effects
can be accounted for by using an effective magnetic field $\MF_\mathrm{eff}(r)$ 
in the calculation of $\Delta N$, given by
\begin{equation}
\MF_\mathrm{eff}(r) = \MF {(1+\mu r)}\exp(-\mu r)
\end{equation}
where $\MF$ is the standard magnetic field, calculated as if there were no paraphotons.

If the source is not pointlike we therefore expect a behavior,
\begin{equation}
\MF_\mathrm{eff}(r) \sim \bigg\{
                           \begin{array}{ll}
                             \MF \exp(-\mu r)&\,\,{\rm{for}} \quad\mu r\gg 1 \\
                             \MF & \,\,{\rm{for}} \quad\mu r\ll 1 \\
                           \end{array},
\label{beff}
\end{equation}
where $r$ is now a typical distance from the source.

For large distances, $r\gg 1/\mu$, we recover the result that the
effective charge vanishes. But for smaller distances, residual effects
of the epsilon charge remain. In typical experiments, the transverse
size\footnote{The important length scale is the distance from the
sources, {\it i.e.}, the currents.}  of the magnetic field is of the order
of $10\,{\rm{cm}}$. Remembering that $1\,{\rm{cm}}\approx
1/(2\times10^{-5}\,{\rm{eV}})$ this can indeed become an important
effect for paraphoton masses of the order of $\mu{\rm{eV}}$.

A similar calculation can be done for the case of two paraphotons. In
this case, the hidden-sector field is not directly coupled to the
electromagnetic field (cf.~Eq.~\eqref{cancel}). The effective epsilon
charge arises, because one paraphoton is massive and the other is
massless, and the cancellation analogous to Fig.~\ref{interaction} is
not complete. Therefore, we are not too surprised by the result,
\begin{equation}
\MF_\mathrm{eff}^\mathrm{MR}(r) = \MF  {\left( 1 -  (1+\mu r)\exp(-\mu r)\right) }.
\end{equation}
For extended sources, we then expect
\begin{equation}
\MF_\mathrm{eff}^\mathrm{MR}(r) \sim \bigg\{
                           \begin{array}{ll}
                             \MF &\,\,{\rm{for}} \quad\mu r\gg 1 \\
                             \MF {\left(0+{\mathcal{O}(\mu r)}\right)}& \,\,{\rm{for}} \quad\mu r\ll 1 \\
                           \end{array},
\end{equation}
where $r$ is again a typical distance from the source.

At large $r\gg 1/\mu$, this indeed looks like a particle with an
effective charge $\epsilon e$. At small distances the charge is
reduced. This effect is exactly the same as the one that is used to
``switch off'' the electric charge of the hidden-sector particle in
astrophysical plasmas in order to avoid the astrophysical bounds on
minicharged particles~\cite{Masso:2006gc}.

\section{Quantitative analysis}
\label{sect:comp}

The contribution of minicharged particles to the rotation and
ellipticity in a pure MCP model has been studied in
Refs.~\cite{Gies:2006ca,Ahlers:2006iz}. If the minicharge originates
from kinetic mixing, the presence of the paraphoton may lead to
significant changes to these signals and will also contribute to LSW
experiments. In this section, we give some explicit examples for the
influence of the paraphoton.

\begin{figure}[p]
\begin{minipage}[c]{\linewidth}\centering
\includegraphics[height=\linewidth]{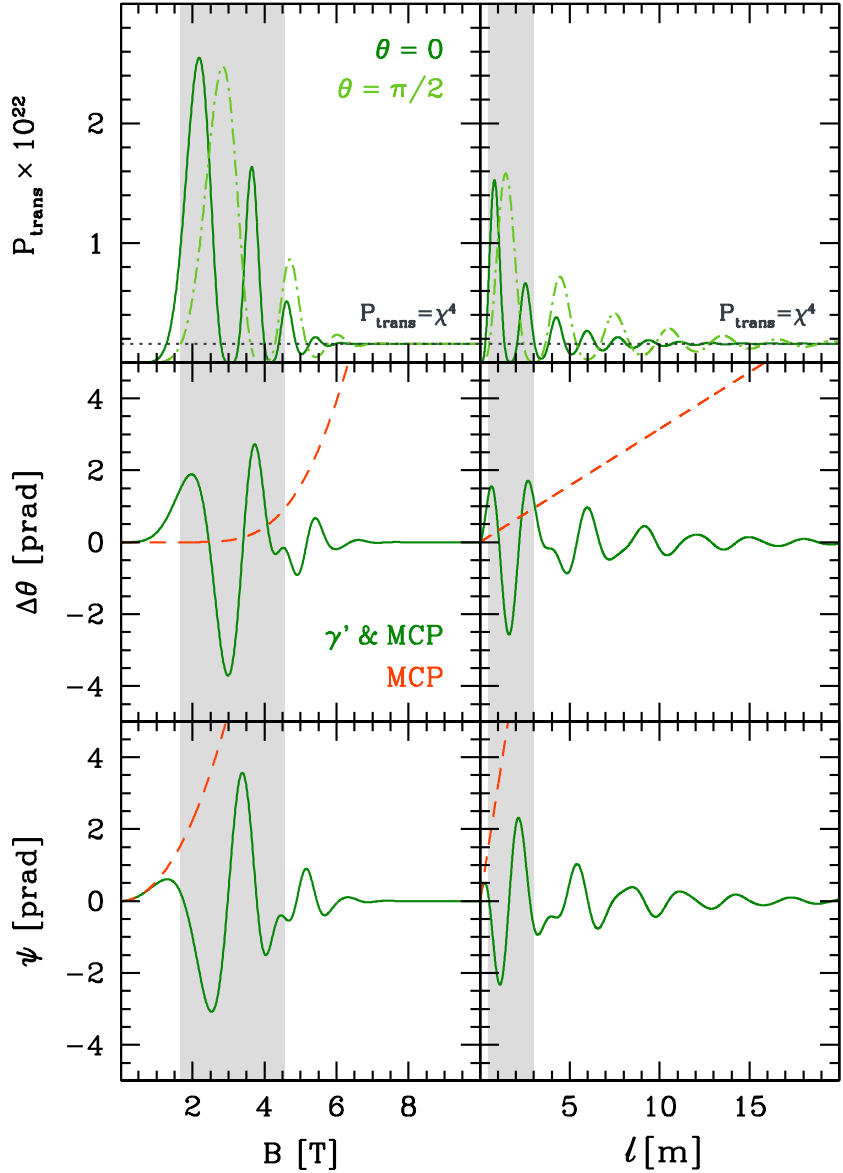}
\end{minipage}
\caption[]{\small Dependence of the regeneration probability
  $P_\text{trans}$ (upper panels), rotation $\Delta \theta$ (center
  panels), and ellipticity $\psi$ (lower panels) on the magnetic filed
  $\MF$ (left panels) and the length $\ell$ of the magnetic region
  inside the cavity (right panels). As a benchmark point we assume one
  massless paraphoton with kinetic mixing parameter
  $\chi=2\times10^{-6}$ and para-coupling $e_h=e$ with a hidden
  Dirac spinor with mass $m_\epsilon=0.1$~eV. The remaining
  experimental parameters are kept at $B=5$~T, $\omega=1$~eV,
  $N_\text{pass}=1$, and $\ell=5$~m in each plot. The photon
  regeneration probability is shown for the case of parallel
  $\theta=0$ (solid line) and orthogonal $\theta=\pi/2$ (dot-dashed
  line) laser polarization. The dotted line indicates the asymptotic
  behavior $P_\text{trans} = \chi^4$. The rotation and ellipticity
  signals assume a polarization of $\theta=\pi/4$. For comparison, the
  dashed line shows the result for rotation and ellipticity without
  massless paraphotons (see Ref.~\cite{Gies:2006ca,Ahlers:2006iz}).  The gray
  shaded band in each plot indicates the oscillation regime,
  corresponding to $\left|K^{\parallel,\perp}\right| < \ell^{-1}$ and
  $\left|\Delta k^{\parallel,\perp}\right| > \ell^{-1}$ (compare
  Eqs.~(\ref{defKD})--(\ref{ellKD})).}\label{expdep}
\end{figure}

Qualitatively, the most obvious difference is the possibility to have
a non-vanishing LSW signal, which is hardly possible without
paraphotons, since the MCPs are unlikely to recombine behind the wall
and produce a photon. The upper panels of Fig.~\ref{expdep} show the
transition probability of photons in a LSW experiment as a function of
the experimental parameters $B$ and $\ell$, the strength and length of
the magnetic field, respectively. Note, that we have non-vanishing
transition probabilities for photons polarized parallelly \emph{and}
perpendicularly to the magnetic field. This in contrast to models with
a single ALP, where the amplitude for the parallel
  (perpendicular) polarization vanishes for a pseudo-scalar (scalar) ALP.

The gray shaded band in the plots indicate a parameter region
  for the experimental setup where the signals have an oscillatory
behavior, corresponding to $\left|K^{\parallel,\perp}\right| <
\ell^{-1}$ and $\left|\Delta k^{\parallel,\perp}\right| > \ell^{-1}$
defined in Eq.~\eqref{defKD}. For $\left|K^{\parallel,\perp}\right|
\gg \ell^{-1}$, the signal becomes constant with
$P_\text{trans}=\chi^4$, whereas it can increase by up to a factor of
16 in the oscillatory region, cf.~Eq.~\eqref{transitionmagnetic}.

The solid lines in the center and lower panels of Fig.~\ref{expdep}
show the rotation and ellipticity of the laser polarization,
respectively in comparison with a pure MCP model (red dashed lines).
In general, the presence of the paraphoton alters the signals
significantly compared to a pure MCP model. In particular, the
seemingly favorable experimental parameters, long and strong magnetic
fields, lead to a small signal. Only inside the oscillatory region the
signals may become comparable to or even larger than the pure MCP
signal, as can be seen from the rotation plots.

\begin{figure}[p]
\begin{minipage}[c]{\linewidth}\centering
\includegraphics[height=\linewidth]{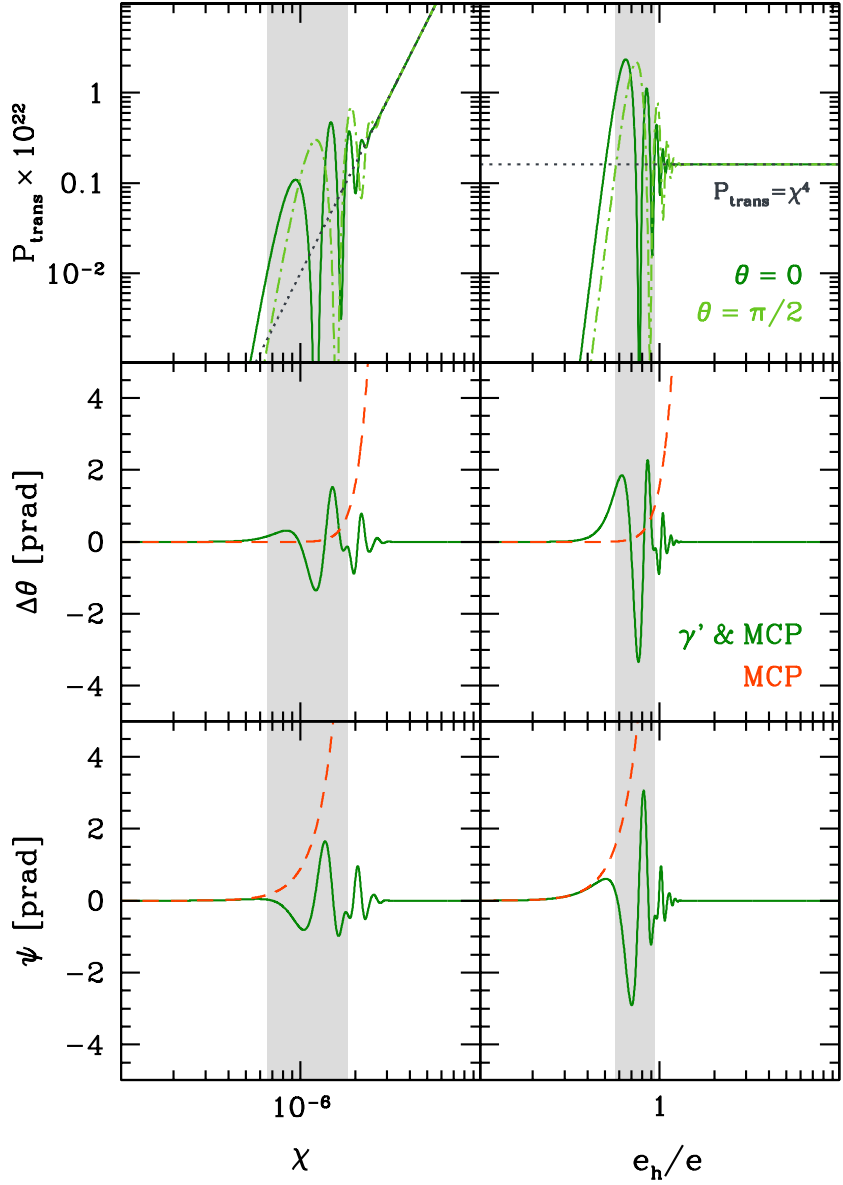}
\end{minipage}
\caption[]{\small Dependence of the regeneration probability $P_\text{trans}$ (upper panels), 
rotation $\Delta \theta$ (center panels), and ellipticity
$\psi$ (lower panels) on the kinetic mixing parameter $\chi$ (left panels), and the relative
para-coupling $e_{h}/e$ (right panels). We use the same benchmark values and notation as in 
Fig.~\ref{expdep}.}\label{moddep}
\end{figure}

\begin{figure}[p]
\begin{minipage}[c]{\linewidth}\centering
\includegraphics[height=\linewidth]{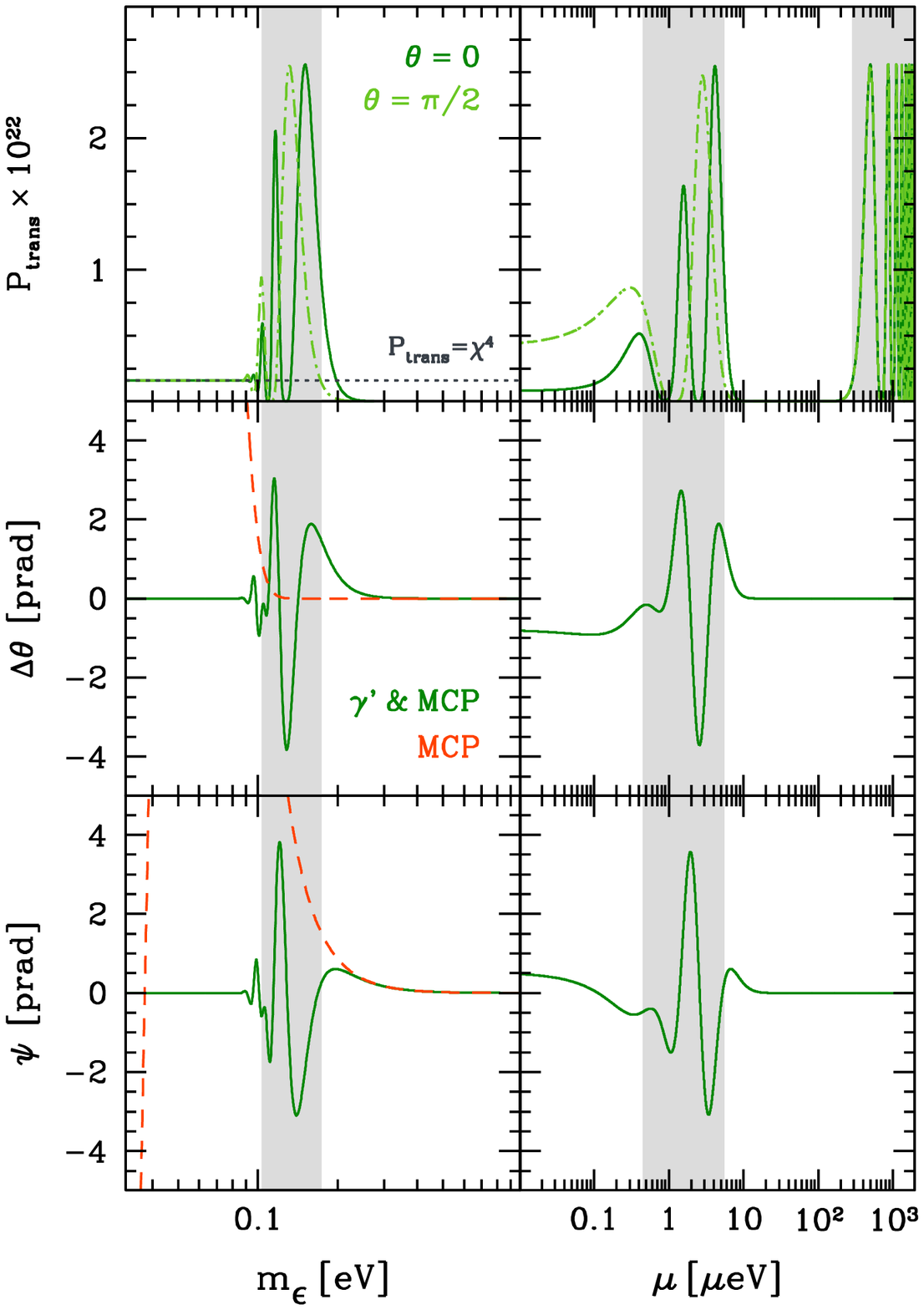}
\end{minipage}
\caption[]{\small Dependence of the regeneration probability
  $P_\text{trans}$ (upper panels), rotation $\Delta \theta$ (center
  panels), and ellipticity $\psi$ (lower panels) on the mass of the
  minicharged particle $m_{\epsilon}$ (left panel) and the mass of the
  paraphoton $\mu$ (right panels).  We use the same benchmark values
  and notation as in Fig.~\ref{expdep}. {In order to calculate the $\mu$
    dependence, we have assumed a typical distance of the laser beam
    from the source of the magnetic field of
    $r=4\,{\rm{cm}}$.}}\label{moddep2}
\end{figure}

Note that these qualitative features are generic to the paraphoton
model, whereas the specific position of the signal peaks depends on
the particular {\it benchmark} point that is used in the plots.  This
becomes apparent in Figs.~\ref{moddep} {and~\ref{moddep2}, where
  now the model parameters are varied while keeping the experimental
  ones fixed. Again, one finds a similar behavior of the signals on
  the kinetic mixing parameter $\chi$, the relative para-coupling
  $e_{h}/e$, the mass of the minicharged particle $m_\epsilon$
  and the mass of the paraphoton $\mu$. For masses of the order of a
  ${\rm{few}}\times 10\,\mu{\rm{eV}}$, the most important effect is the
  reduction of the effective magnetic field as discussed in
  Sect.~\ref{finite}, since masses in the $\mu{\rm{eV}}$ range
  are not big enough to lead to a sizeable
  transition probability from oscillations due to the mass alone. For
  bigger masses $\gtrsim {\rm{meV}}$, the photon-paraphoton
  oscillations are driven by the mass term. In this region, the signal
  does not change if the magnetic field is switched off.}

The reason for the fact that ellipticity and rotation become
  insensitive to the model parameters for large magnetic field length
  or strength can easily be understood heuristically: owing to the
  nonzero depletion coefficient $\kappa$ for the photon interaction
  state, the combined photon-paraphoton state evolves nonunitarily
  over long distances into that mixed state which does not interact
  with the hidden fermions $h$. For this state, the effective
  refractive index and depletion coefficient {approach the
    trivial vacuum values}; consequently,
  any further ellipticity or rotation effects are absent in this
  regime.

It is interesting to observe that the ellipticities in the
  paraphoton model deviate from the pure MCP {model towards}
  smaller values in the oscillatory region, whereas the rotations also
  exhibit peaks that exceed the pure MCP value, see, {\it e.g.},
  Figs.~\ref{expdep} and \ref{moddep}. The reason for these pronounced
  rotation peaks in the paraphoton model lies in a nontrivial
  interplay between the paraphoton and the minicharged fluctuations,
  as is visible from the second term in Eq.~\eqref{rotKD}. In pure MCP
  models, rotation is induced by photon loss due to MCP production
  ({first} term in Eq.~\eqref{rotKD}) for which mass-threshold and
  phase-space conditions have to be satisfied. With a light
  paraphoton, these conditions are much more relaxed; for instance, a
  photon-paraphoton transition via a virtual intermediate MCP state
  can be possible even if the photon energy is too small to excite a
  real MCP pair. This rotation-inducing effect is a genuine feature of
  models with both MCPs and paraphotons. The model-parameter range where
  these rotation peaks appear is also a promising candidate for
  parameterizing the anomalous PVLAS rotation signal
  \cite{Zavattini:2005tm}; a precise fit to the corresponding allowed
  parameter range, however, is beyond the scope of the present work.

\begin{figure}[t]
\begin{minipage}[t]{\linewidth}\centering
\includegraphics[height=0.5\linewidth]{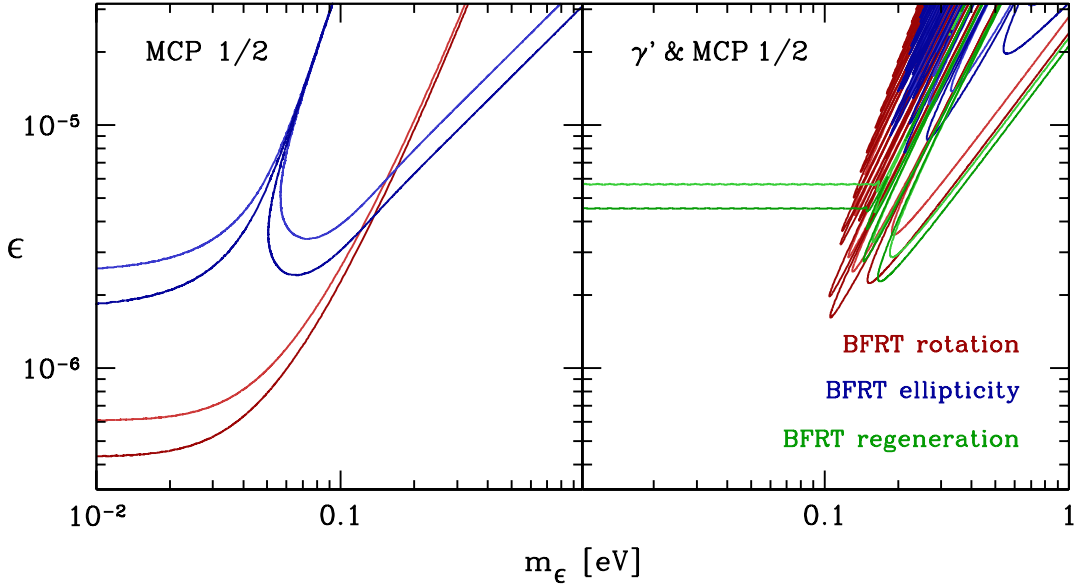}
\end{minipage}
\caption[]{\small{ Exclusion limits from the BFRT experiment. 
The dark (light) contours show the 2$\sigma$ (5$\sigma$)
    exclusion limits of charge $\epsilon$ and mass
  $m_\epsilon$ of a Dirac spinor corresponding to the measurements of the BFRT collaboration. (For simplicity, we assume a
  constant magnetic field amplitude of $B=2$~T for the calculation of
  the rotation and ellipticity signal.) The left panel shows the excluded region in the pure MCP model. The right panel shows the
  results including a massless paraphoton with para-coupling
  $e_{h}=e$. The loss of sensitivity for small masses is
  partially compensated by the results of the photon regeneration
  experiment.}}\label{limits}
\end{figure}

The BFRT collaboration~\cite{Cameron:1993mr,Ruoso:1992nx} performed a
pioneering experiment searching for the rotation, ellipticity, and
photon regeneration signals. From the non-observation of a signal
one can infer exclusion regions for the MCP scenario as well as
extensions with paraphotons. The left plot of Fig.~\ref{limits} shows
the excluded region of mass $m_\epsilon$ and charge $\epsilon$ in the
pure MCP model. In this case, the model is not constrained by the
regeneration measurement.

This is different for paraphoton models, as can be seen in the right
plot of the same figure. For small masses, rotation and ellipticity
do not represent sensitive probes of the model parameter space.
However, the regeneration limit puts a constant upper bound on the
charge $\epsilon$ at small masses $m_\epsilon$ and not too small
$e_{h}$, corresponding to the asymptotic behavior of the transition
probability $P_\text{trans}\rightarrow\chi^4$. This partially
compensates for the loss of sensitivity of the optical measurements.
This demonstrates that LSW experiments are complementary to
polarization measurements.

\begin{figure}[t]
\begin{minipage}[t]{\linewidth}\centering
\includegraphics[width=\linewidth]{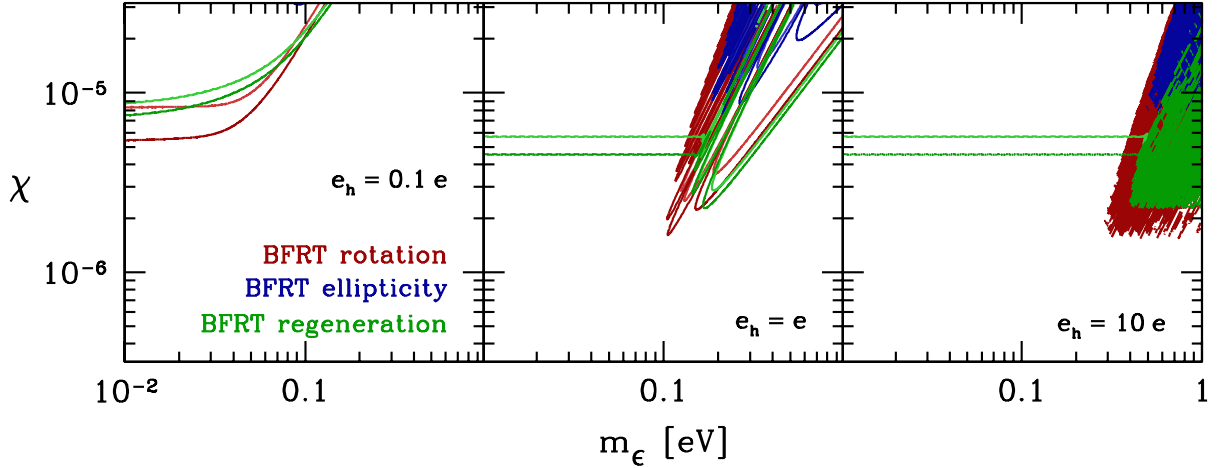}
\end{minipage}
\caption[]{\small {The limit from BFRT
measurements on the
  kinetic mixing parameter for various {values} of the para-coupling
  $e_h$. The 2$\sigma$ (5$\sigma$)
    exclusion limits are plotted as dark (light) contours.}
}\label{limits_reg}
\end{figure}

Of course, the results are also somewhat dependent on the gauge coupling of 
the paraphoton, $e_h$. But, as can be seen from Fig.~\ref{limits_reg}, even a variation
of the gauge coupling by one order of magnitude around the natural value $e$ 
leads to relatively small changes in the limit on $\chi$ obtainable from the BFRT 
regeneration data. This is a significant advantage of LSW experiments.   

The qualitative dependence of the limits from LSW measurements on the remaining model parameter, 
the paraphoton mass $\mu$,  
can already be inferred from the right uppermost panel in Fig.~\ref{moddep2}. 
If we assume a typical distance of the laser beam from the source of the
magnetic field of the order of 5~cm, photon regeneration is sensitive 
in the range $\mu\lesssim 10$~$\mu$eV to oscillations induced by the
magnetic field. For bigger masses, this effect is extremely suppressed because
the magnetic field is effectively zero, as can be seen from Eq.~\eqref{beff}. 
The signal is then driven by oscillations via the  mass term, and the
BFRT bounds are as in Fig.~\ref{fig1}. 

\begin{figure}[t]
\begin{minipage}[t]{\linewidth}\centering
\includegraphics[height=0.39\linewidth]{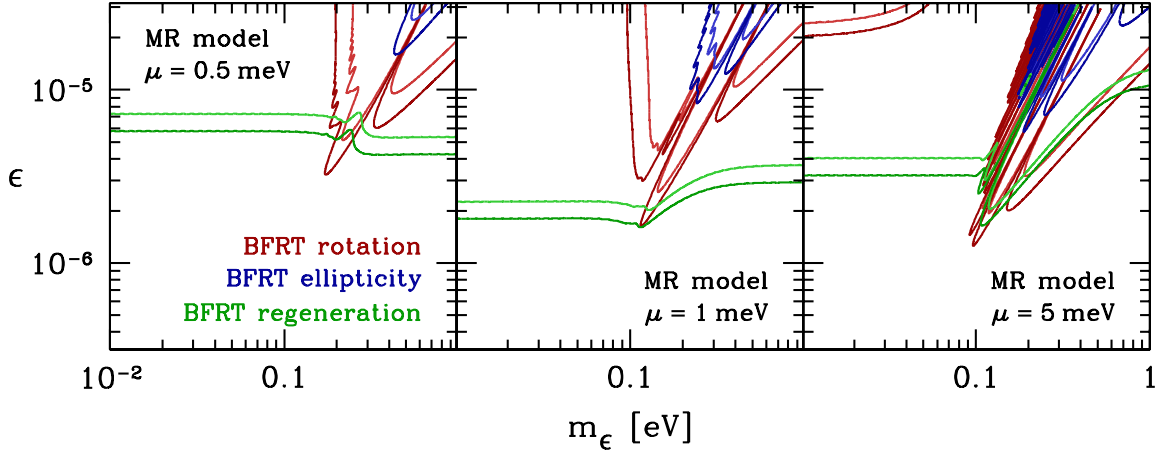} 
\end{minipage}
\caption[]{\small {The limit from BFRT
measurements on
    the kinetic mixing parameter in the two-paraphoton model presented
    in Ref.~\cite{Masso:2006gc} ($e_h=e$). The 2$\sigma$ (5$\sigma$)
    exclusion limits are plotted as dark (light) contours. (The chosen
    values $\mu\sim\,{\rm{meV}}$ are small enough to avoid the
    astrophysical constraints and big enough such that effects of the
    finite size of the magnetic field play no
    role.)}}\label{limits_MR_1}
\end{figure}

Finally, let us comment on the two-paraphoton model of Ref.~\cite{Masso:2006gc}. 
In this model, regeneration again leads to the best bounds, as can be seen 
from Fig.~\ref{limits_MR_1}. 

\section{Conclusions}\label{conclusions}

{Constraining the multitudinous possibilities to
extend the standard model} of particle physics requires powerful
laboratory tools that do not only search for new particles at higher
and higher masses, but also for weakly coupled hidden sectors with
potentially light particles. In this work, we have shown  that
light-shining-through-walls (LSW) experiments represent one of these
desired tools to specifically look for a hidden sector with additional
U(1) paraphoton gauge groups -- in addition to their discovery
potential of axion-like particles (ALP), as conventionally discussed
in the literature.  This becomes evident from Figs.~\ref{fig1}, 
\ref{limits}, and \ref{limits_MR_1}, in which we present limits obtained
from the BFRT LSW experiment.

Owing to their specific dependence on both model as well as
experimental parameters  (see Figs.~\ref{expdep},
\ref{moddep}, and \ref{moddep2}), LSW experiments are
also ideally suited to distinguish between different models such as
those involving ALPs or paraphotons.  One important
example of a feature that allows to distinguish between ALPs and
paraphotons is the dependence on the polarization of the laser beam.
For ALPs, we expect a signal only for one polarization, parallel {\em
  or} perpendicular to the magnetic field. In paraphoton models, we
expect an LSW signal for {\em both} polarizations. {Also, the
  dependence of the regeneration rates on experimental parameters such
  as the laser frequency and the magnetic field are different for the
  different models and thus provide for further decisive
  distinguishing criteria.}

Polarization experiments provide complementary information (cf.~Fig.~\ref{limits}). 
They are
especially sensitive to pure minicharged particle models for which no
signal is expected in LSW experiments. However, in paraphoton models, 
their sensitivity is limited.  

In conclusion, regenerating (para-)light from the hidden sector allows to test a large
class of natural extensions of the standard model.

The discovery potential of optical experiments for features of the
hidden sector is certainly not exhausted by our present study.  For
instance, the use of the rapidly evolving pulsed high-intensity laser
systems for this type of fundamental physics challenges needs to be
explored much further, see, {\it e.g.},
\cite{Heinzl:2006xc,DiPiazza:2006pr}. Also, nonlinear collective
effects in photon-plasma interactions \cite{Marklund:2006my} may serve
as an amplifier of signatures of the hidden sector.  Finally,
experiments with large electric fields, where light minicharged
particles could be produced by the Schwinger mechanism, can provide
additional insights~\cite{Gies:2006hv}.

\section*{Acknowledgements}

HG acknowledges support by the DFG under contract Gi 328/1-4
(Emmy-Noether program). JR acknowledges enlightning conversations with Eduard Mass\'o, Jhon
Calsamiglia and Javier Virto, as well as support by a contract from the
Universidad Aut\'oma de Barcelona, by the CICYT Project FPA2005-05904 and
the DURSI Project  2005SGR00916.

\begin{appendix}
\section{Refractive index for photons in a magnetic field}\label{refraction}
The loop diagram depicted in Fig.~\ref{convertAA} gives the
contribution of hidden-sector particles to the complex refractive
index for photons.  The value of this diagram is well known
\cite{Tsai:1975iz}. Let us define
\begin{equation}
\epsilon^{2} e^{2}\Delta N=\epsilon^{2} e^{2}(\Delta n+\frac{1}{{2\omega}}\ii
\kappa)=n-1.
\end{equation}
The contribution from intermediate Dirac spinors (``Dsp'') and scalars
(``sc'') with an effective coupling $\epsilon e$ to photons is given
as
\begin{equation}
\Delta n_{\parallel,\perp}^{\text{Dsp}/\text{sc}}(\epsilon e\MF,m_{\epsilon})=
-\frac{1}{16\pi^2}\left(\frac{\epsilon\,e\MF}{m^{2}_{\epsilon}}\right)^{2}
I_{\parallel,\perp}^{\text{Dsp}/\text{sc}}(\chiS),
\end{equation}
with
\begin{eqnarray}
I_{\parallel,\perp}^{\rm Dsp}(\chiS)\!\!&=&\!\!2^{\frac{1}{3}}\left(\frac{3}{\chiS}\right)^{\frac{4}{3}}
\int^{1}_{0} {\rm d}v\,
\frac{\left[\left(1-\frac{v^2}{3}\right)_{\parallel},
\left(\frac{1}{2}+\frac{v^2}{6}\right)_{\perp}\right]}{(1-v^{2})^{\frac{1}{3}}}
\times\tilde{e}^{\prime}_{0}\left[\begin{scriptstyle}-
\left(\frac{6}{\chiS}\frac{1}{1-v^2}\right)^{\frac{2}{3}}\end{scriptstyle}\right]
\\\nonumber
 &&\quad\quad\quad\quad\quad\quad\quad\quad\quad\quad\quad=
 \begin{cases}
-\frac{1}{45} \left[(14)_\parallel,(8)_\perp\right], & \text{for}\,\,\chiS\ll 1\text{,} \\
\frac{9}{7}\frac{\pi^{\frac{1}{2}}2^{\frac{1}{3}}
\left(\Gamma(\left(\frac{2}{3}\right)\right)^{2}}{\Gamma\left(\frac{1}{6}\right)}
\chiS^{-4/3}\left[ (3)_\parallel,(2)_\perp\right],& \text{for}\,\,\chiS\gg 1\text{.}
\end{cases}.
\end{eqnarray}
and
\begin{eqnarray}
I_{\parallel,\perp}^{\text{sc}}(\chiS)
\!\!&=&\!\!\frac{2^{\frac{1}{3}}}{2}\left(\frac{3}{\chiS}\right)^{\frac{4}{3}}
\int^{1}_{0} {\rm d}v\,
\frac{\left[\left(\frac{v^2}{3}\right)_{\parallel},
\left(\frac{1}{2}-\frac{v^2}{6}\right)_{\perp}\right]}
{(1-v^{2})^{\frac{1}{3}}}
{\times\tilde{e}^{\prime}_{0}
\left[
  \begin{scriptstyle}-
    \left(\frac{6}{\chiS}\frac{1}{1-v^2}\right)^{\frac{2}{3}}
  \end{scriptstyle}
\right]}
\\\nonumber
 &&\quad\quad\quad\quad\quad\quad\quad\quad\quad\quad\quad=
   \begin{cases}  -
    \frac{1}{90} \left[(1)_\parallel,(7)_\perp\right],
    & \text{for} \,\,\chiS\ll 1\,\,\text{,}\\
    \frac{9}{14}\frac{\pi^{\frac{1}{2}}2^{\frac{1}{3}}
      \left(\Gamma\left(\frac{2}{3}\right)\right)^{2}}
    {\Gamma\left(\frac{1}{6}\right)}
    \chiS^{-4/3}\left[
    (\frac{1}{2})_\parallel,(\frac{3}{2})_\perp\right],& \text{for}
    \,\,\chiS\gg 1\,\,\text{.}
  \end{cases}
\end{eqnarray}
Here, the dimensionless parameter $\chiS$ is defined as
\begin{equation}
\label{chi}
\chiS \equiv  \frac{3}{2} \frac{\omega}{m_\epsilon} \frac{\epsilon e \MF}{m_\epsilon^2}
= 88.6\ \epsilon\ \frac{\omega}{m_\epsilon}\
\left( \frac{\rm eV}{m_\epsilon}\right)^2
\left( \frac{\MF}{\rm T}\right)
\,.
\end{equation}
The symbol $\tilde{e}_{0}$ denotes the generalized Airy function,
\begin{equation}
\tilde{e}_{0}(t)=\int^{\infty}_{0}{\rm d}x\,\sin\left(tx-\frac{x^3}{3}\right),
\end{equation}
and $\tilde{e}^{\prime}_{0}(t)={\rm{d}}\tilde{e}_{0}(t)/{\rm{d}}t$.

Similarly,
\begin{eqnarray}\label{absorption}
{\kappa_{\parallel,\perp}^{\rm Dsp}(\epsilon e\MF,m_{\epsilon}) \ell}
\!\!&=&\!\! \frac{1}{2}\epsilon \frac{e}{4\pi} \frac{\MF \ell }{m_\epsilon}\,
T_{\parallel,\perp}^{\rm Dsp}(\chiS )
,
\end{eqnarray}
$T_{\parallel,\perp }(\chiS )$ has the form of a parametric
integral~\cite{Tsai:1974fa},
\begin{eqnarray}
\label{absorb}
&&\!\!\!\!\!\!T_{\parallel,\perp}^{\rm Dsp} =
\frac{4\sqrt{3}}{\pi\chiS}
\int\limits_0^1 {\rm d}v\
K_{2/3}\left( \frac{4}{\chiS}\frac{1}{1-v^2}\right)
\times
\frac{\left[ \left( 1-\frac{1}{3}v^2\right)_\parallel,
\left(\frac{1}{2} +\frac{1}{6}v^2\right)_\perp
\right]}{(1-v^2)}
\\\nonumber
&&
\quad\quad\quad\quad\quad\quad\quad\quad\quad\quad\quad\quad\quad\quad\quad\quad
= \begin{cases}
\sqrt{\frac{3}{2}}\ {\rm e}^{-4/\chiS}\ \left[(\frac{1}{2})_\parallel,(\frac{1}{4})_\perp\right], & \text{for}
\,\,\chiS\ll 1\,\,\text{,} \\
\frac{2\pi}{\Gamma\left(\frac{1}{6}\right)\Gamma\left(\frac{13}{6}\right)}
\chiS^{-1/3}\left[ (1)_\parallel,(\frac{2}{3})_\perp\right],& \text{for}  \,\,\chiS\gg 1\,\,\text{,}
\end{cases}
\end{eqnarray}
and
\begin{eqnarray}
\label{eqDB4}
&&\!\!\!\!\!\!T_{\parallel,\perp}^{\text{sc}} =
\frac{2\sqrt{3}}{\pi\chiS}
\int\limits_0^1 {\rm d}v\
K_{2/3}\left( \frac{4}{\chiS}\frac{1}{1-v^2}\right)
\times
\frac{\left[ \left(\frac{1}{3}v^2\right)_\parallel,
\left(\frac{1}{2} -\frac{1}{6}v^2\right)_\perp
\right]}{(1-v^2)}
\\\nonumber
&&
\quad\quad\quad\quad\quad\quad\quad\quad\quad\quad\quad\quad\quad\quad\quad\quad
=\begin{cases}
\frac{1}{2} \sqrt{\frac{3}{2}}\ {\rm e}^{-4/\chiS}\
\left[(0 )_\parallel,(\frac{1}{4})_\perp\right], & \text{for}
\,\,\chiS\ll 1\,\,\text{,} \\
\frac{\pi}{\Gamma\left(\frac{1}{6}\right)\Gamma\left(\frac{13}{6}\right)}
\chiS^{-1/3}\left[ (\frac{1}{6})_\parallel,(\frac{1}{2})_\perp\right],&
 \text{for} \,\,\chiS\gg 1\,\,\text{.}
\end{cases}
\end{eqnarray}

These expressions have been derived to leading order in an expansion
for high frequency
{\cite{Toll:1952rq,Klepikov:1954,Erber:1966vv,Baier:1967,Klein:1968,Dittrich:2000zu}},
\begin{equation}
\label{semiclhf}
\frac{\omega}{2m_\epsilon}\gg  1,
\end{equation}
and for a high number of allowed Landau levels of the minicharged
particles \cite{Daugherty:1984tr},
\begin{eqnarray}
\nonumber
\Delta N_{\rm{p}}\!\!&=&\!\!\frac{\Delta N_{\rm{Landau}}}{2}=
\frac{1}{12}\left(\frac{\omega^{2}}{\epsilon\,e\MF}\right)^{2}
\left( \frac{\Delta\omega}{\omega}+ \frac{\Delta \MF}{2 \MF}\right) \gg 1
\\[1.5ex]\label{peaks}
&&\quad\quad\quad\quad\Leftrightarrow \epsilon\ll  4.9\times 10^{-3} \left(\frac{\omega}{\rm{eV}}\right)^{2}
\left(\frac{\rm{T}}{\MF}\right)
\left(\frac{\Delta\omega}{\omega} + \frac{\Delta \MF}{2 \MF}\right)^{\frac{1}{2}}.
\end{eqnarray}
{In realistic experiments, the
  variation} $\Delta\omega/\omega$ is typically small compared to
$\Delta \MF/\MF\gtrsim 10^{-4}$.

\section{Preparation of the initial state and cavity
  effects}\label{initialstate}

Let us devote a few thoughts to the preparation of the initial state.
In Sects. \ref{lsw0} and \ref{lsw1}, we always started with a pure
photon interaction state, $(A,B)=(1,0)$ or $(A,B_{1},B_{2})=(1,0,0)$.
Is this the correct state for a realistic experiment? Naively, the
answer is yes, because the light is produced by ordinary matter which
interacts only with the photon interaction eigenstate. {Still, one
  might wonder whether the laser apparatus might be so precise that it
  can prepare eigenstates of the energy and the momentum
  simultaneously. }

\begin{figure}[t]
\begin{center}
\includegraphics[width=0.55\linewidth]{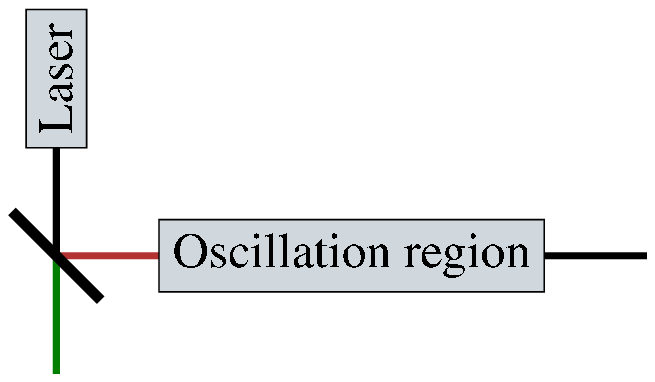}
\end{center}
\caption{\small Sketch of the initial-state preparation in a
  photon-paraphoton oscillation experiment. The laser (including its
  optical elements) produces some unknown mixture of photon and
  paraphoton (black line). Now, this beam is redirected via a mirror
  (black diagonal) into the oscillation region. However, the mirror
  interacts only with the interaction eigenstate of the photon (red).
  The paraphoton interaction state simply passes through the mirror
  (green).  Therefore, we have a pure photon interaction state at the
  beginning of the oscillation region. If the photon interaction state
  does not coincide with the propagation eigenstates, {\it i.e.}, if we have
  mixing, we have a mixed interaction state (black) at the end of the
  oscillation region.}\label{mirror}
\end{figure}

Figure \ref{mirror} shows why this is not  {really relevant for}
the case of a typical setup where the laser beam is coupled into the
oscillation region via a mirror (we believe that in most experiments
such a redirection of the beam is employed at some stage of the
experiment; in BFRT as well as PVLAS this is indeed the case).  It is
simply the mirror that again selects the interaction state and directs
only the photon interaction state into the right direction towards the
oscillation region. The paraphoton interaction state simply passes
through the mirror and is {lost}.

The bottom line is that the last mirror that couples the beam into the
oscillation region selects a pure photon interaction state, and this
determines the initial condition.

Next, we address the question as to whether some optical elements as,
  {\it e.g.}, a Fabry-Perot cavity with a high finesse could again select a
  momentum eigenstate. If so, such a state would have a well-defined wavelength and
  would therefore correspond to a propagation eigenstate -- destroying
  possible oscillations.

\begin{figure}[t]
\begin{center}
\includegraphics[width=0.4\linewidth]{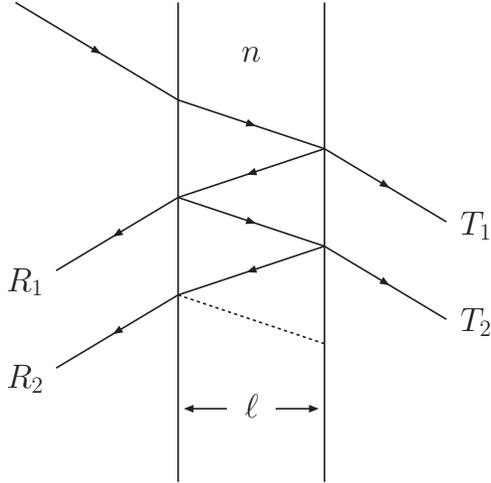}
\end{center}
\caption{\small Light path inside a Fabry-Perot cavity.}
\label{fabry}
\end{figure}

In ordinary optics, the transmission coefficient for a Fabry-Perot cavity is 
\begin{equation}
\label{fpwithout}
T_{\text{FP}}=\frac{T^2}{1+R^2-2R\cos(\delta)}, 
\end{equation}
with
\begin{equation}
\delta=2 k \ell \cos(\theta).
\end{equation}
Here, $R$ and $T$ are the transmission and reflection coefficients of the mirrors. 
We assume no absorption, {\it i.e.}~$T=1-R$.
The transmission is strongly peaked around $\delta=0$ and effectively filters out a 
very narrow wavelength interval
of width
\begin{equation}
\frac{\delta\lambda}{\lambda}=\frac{\lambda}{2{\mathcal{F}}\ell\cos(\theta)},
\end{equation}
where $\ell$ is the length of the cavity, and
\begin{equation}
{\mathcal{F}}=\frac{\pi}{2\arcsin\left(\frac{1-R}{2\sqrt{R}}\right)}\approx \frac{2\pi}{1-R}
\end{equation}
denotes its finesse; the approximation in the last step holds for $1-R\ll 1$.
$\theta$ is the angle of the incident light (cf.~Fig.~\ref{fabry}) which
we will take to be $\theta=0$ for simplicity.

We can now study what happens in a model with a paraphoton. We start
with a pure photon interaction state $(A,B)=(1,0)$ at the entrance to
the cavity.  Using the propagation eigenstates found in Sect.
\ref{lsw0}, we find the amplitude after the first pass through the
cavity,
\begin{equation}
T_{1}=\left(
       \begin{array}{c}
         A_{1} \\
         B_{1} \\
       \end{array}
     \right)
=T \exp(\ii k\ell)\left(
                   \begin{array}{c}
                     \frac{1+\chi^2\exp(-\Delta k \ell)}{1+\chi^2} \\
                     \frac{\chi(1-\exp(-\Delta k \ell))}{1+\chi^2} \\
                   \end{array}
                 \right).
\end{equation}
Taking into account that only the photons and not the paraphotons are
reflected by the mirrors, we can easily find also the amplitude
for the second transmitted beam,
\begin{equation}
T_{2}=\left(
       \begin{array}{c}
         A_{2} \\
         B_{2} \\
       \end{array}
     \right)
=T \exp(3\ii k\ell)R \left(\frac{1+\chi^2\exp(-\Delta k \ell)}{1+\chi^2}\right)^{2}\left(
                   \begin{array}{c}
                     \frac{1+\chi^2\exp(-\Delta k \ell)}{1+\chi^2} \\
                     \frac{\chi(1-\exp(-\Delta k \ell))}{1+\chi^2} \\
                   \end{array}
                 \right).
\end{equation}

Resumming $A_{\rm{trans}}=A_{1}+A_{2}+\ldots$, we find
the total transition coefficient for the Fabry-Perot cavity,
\begin{equation}
\hat{T}_{\rm{FP}}=|A_{\rm{trans}}|^{2}=\frac{|TM|^{2}}{1+|M^{2}R|^{2}-2|M^{2}R|\cos(\delta+\alpha)}, 
\end{equation}
where
\begin{equation}
M=\frac{1+\chi^2\exp(-\Delta k\ell)}{1+\chi^2}=:|M|\exp(\ii \alpha)
\end{equation}
is the photon-to-photon amplitude for one pass through the cavity.
For small $\chi$, we find
\begin{equation}
|M|=1-4\chi^2\sin^{2}\left(\frac{\Delta k\ell}{2}\right), \quad \alpha=2\chi^2\sin(\Delta k\ell).
\end{equation}
If $\chi^2\ll 1/{\mathcal{F}}$, {\it i.e.}~if more photons escape from the
cavity via transmission than via conversion into paraphotons, $\alpha$
and $|M|-1$ provide only small corrections to the result without
paraphotons \eqref{fpwithout}, and the cavity selects essentially the same
wavelengths around $\delta\approx 2\pi n$ as without paraphotons.  For
example, in the PVLAS experiment with ${\mathcal{F}}\sim 10^5$ this
condition is easily fulfilled for $\chi \lesssim 10^{-5}$.

Let us now turn to the paraphotons exiting the cavity. The
transmission coefficient for paraphotons, or, in other words, the
photon-paraphoton conversion probability, is (for small $\chi$),
\begin{eqnarray}
T_{\rm{para}}\!\!&=&\!\!|B_{\rm{trans}}|^{2}=4 |T|\chi^2\sin^{2}\left(\frac{\Delta k\ell}{2}\right)
\frac{1}{1+|M^{2}R|^{2}-2|M^{2}R|\cos(\delta+\alpha)}
\\\nonumber
\!\!&\approx&\!\!\frac{2{\mathcal{F}}}{\pi}\chi^{2}\sin^{2}\left(\frac{\Delta k\ell}{2}\right)
\\\nonumber
\!\!&\approx&\!\! \frac{N_{\rm{pass}}+1}{2}\chi^{2}\sin^{2}\left(\frac{\Delta k\ell}{2}\right).
\end{eqnarray}
The last two lines hold for $\chi^2\ll 1/{\mathcal{F}}$ and $\delta
\approx 2\pi n$, {\it i.e.}, for incident photons in resonance with the
cavity.

To summarize, as long as paraphotons are a ``small'' effect we rediscover
the naively expected result.

\section{Regeneration probability for the two paraphoton model}\label{MRregen}
In this appendix, we give the explicit formulas for the regeneration probability, rotation and ellipticity 
in the model of Ref.~\cite{Masso:2006gc}
with two paraphotons.

With the abbreviation $\sqrt{(e^{2}_{{h}}\Delta N)^2+(\mu^2/4\omega^2)^2}=\alpha+\ii\beta$ and 
$\Delta N = \Delta n + \ii\kappa/2\omega$, we define the inverse oscillation and absorption lengths 
$\Delta_\pm$ and $\kappa_\pm$, respectively, as
\begin{align*}
\Delta_\pm &=\omega \,e^{2}_{{h}}\Delta n-\frac{\mu^2}{4\omega} \mp \omega,\,\alpha&\kappa_\pm&=
e^{2}_{{h}}\kappa\mp2\omega\,\beta .
\end{align*}
The transition probability ($\ell_1=\ell_2=\ell$ and $N_{\rm{pass}}=1$) is given by the squared sum of the 
amplitudes for the transition of the wall through the two different paraphotons
\begin{equation}
P_{\rm{trans}} = \left| {\cal
    A}_{\gamma\rightarrow\gamma_1' \rightarrow\gamma}
+{\cal A}_{\gamma\rightarrow\gamma_2'\rightarrow\gamma}\right|^2\,,
\end{equation}
that can be expressed as
\begin{equation}\label{P2}
P_{\rm{trans}} = \frac{\chi^4}{X_0^2+Y_0^2}\left[\left(2X_0
    +X_+C_++X_-C_-+Y_+S_++Y_-S_-\right)^2
+\left(\begin{matrix}X_{{i}}\rightarrow Y_{{i}}\\Y_{{i}}\rightarrow-X_{{i}}\end{matrix}\right)\right]. 
\end{equation}
Here, {the index $i$ denotes $i=(+,-,0)$,} and we have used the functions
\begin{align*}
S_\pm(\ell) &= \exp\left(-\kappa_\pm \ell\right)\sin(2\Delta_\pm \ell)-
2\exp\left(-\kappa_\pm \ell/2\right)\sin(\Delta_\pm \ell)\,,\\
C_\pm(\ell) &= \exp\left(-\kappa_\pm \ell\right)\cos(2\Delta_\pm \ell)-
2\exp\left(-\kappa_\pm \ell/2\right)\cos(\Delta_\pm \ell)\,,
\end{align*}
with coefficients
\begin{eqnarray}
X_\pm \!\!&=&\!\! 16\left(2e^{2}_{{h}}\Delta n \frac{e^{2}_{{h}}\kappa}{2\omega}\pm\beta e^{2}_{{h}}\Delta n
\pm \alpha \frac{e^{2}_{{h}}\kappa}{2\omega}\right)\,,
\\\nonumber
Y_\pm \!\!&=&\!\! \frac{\mu^4}{\omega^4}+16 e^{2}_{{h}}\Delta n (e^{2}_{{h}}\Delta n \pm \alpha)
- 16\frac{e^{2}_{{h}}\kappa}{2\omega}\left(\frac{e^{2}_{{h}}\kappa}{2\omega}\pm\beta\right)\,,
\\\nonumber
X_0 \!\!&=&\!\! 32\,e^{2}_{{h}}\Delta n\frac{e^{2}_{{h}}\kappa}{2\omega}\,,
\\\nonumber
Y_0 \!\!&=&\!\! \frac{\mu^4}{\omega^4}+16\left((e^{2}_{{h}}\Delta n)^2 - 
\left(\frac{e^{2}_{{h}}\kappa}{2\omega}\right)^2\right)\,.
\end{eqnarray}
The case $\MF=0$ corresponds to $\Delta n=\kappa=0$, giving $\alpha=\frac{\mu^2}{4\omega^2}$ and $\beta=0$,  
as well as
$\kappa_\pm = \Delta_- = 0$ and $\Delta_+=-\frac{\mu^2}{2\omega}$. It is straightforward to check that in 
this case Eq.~(\ref{P2})
reduces to our previous result Eq.~(\ref{P1}) with $N_\text{pass}=1$ and $\ell_1=\ell_2$.

For the photon to photon amplitude we find,
\begin{eqnarray}
\!\!{\rm{Re}}(A_{\gamma\to\gamma})\!\!&=&\!\!1-2\chi^2+\frac{\chi^2}{\alpha^{2}+\beta^{2}}
\\\nonumber
&&\!\!\!\!\!\!\!\!\!\!\!\!\!\!\!\!\!\!\!\!\!\!\!\times\bigg[Z_{+}
\cos(\Delta_{+}\ell)\exp(-\kappa_{+}\ell/2)+(+\rightarrow -)
+Z_{0}\sin(\Delta_{+}\ell)\exp(-\kappa_{+}\ell/2)-(+\rightarrow -)
\bigg]
\end{eqnarray}
and
\begin{eqnarray}
\!\!{\rm{Im}}(A_{\gamma\to\gamma})\!\!&=&\!\!\frac{\chi^2}{\alpha^{2}+\beta^{2}}
\\\nonumber
&&\!\!\!\!\!\!\!\!\!\!\!\!\!\!\!\!\!\!\!\!\!\!\!\times\bigg[Z_{+}
\sin(\Delta_{+}\ell)\exp(-\kappa_{+}\ell/2)+(+\rightarrow -)
+Z_{0}\cos(\Delta_{-}\ell)\exp(-\kappa_{-}\ell/2)-(-\rightarrow +)
\bigg],
\end{eqnarray}
where 
\begin{eqnarray}
Z_{\pm}\!\!&=&\!\!\alpha\left(\alpha\pm e^{2}_{h}\Delta n\right)+\beta\left(\beta\pm 
\frac{e^{2}_{h}\kappa}{2\omega}\right),
\\\nonumber
Z_{0}\!\!&=&\!\!\beta e^{2}_{h}\Delta n-\alpha\frac{e^{2}_{h}\kappa}{2\omega}.
\end{eqnarray}
It is now straightforward to insert this into Eqs.~\eqref{rotation} and \eqref{ellipticity} 
to obtain the rotation and ellipticity, respectively.

\end{appendix}


\begin{thebibliography}{10}

\bibitem{Cameron:1993mr}
  R.~Cameron {\it et al.} [BFRT Collaboration],
  Phys.\ Rev.\  D {\bf 47} (1993) 3707.

\bibitem{Zavattini:2005tm}
  E.~Zavattini {\it et al.}  [PVLAS Collaboration],
  Phys.\ Rev.\ Lett.\  {\bf 96} (2006) 110406
  [arXiv:hep-ex/0507107].

\bibitem{Chen:2006cd}
  S.~J.~Chen, H.~H.~Mei and W.~T.~Ni [Q\&A Collaboration],
arXiv:hep-ex/0611050.

\bibitem{Maiani:1986md}
  L.~Maiani, R.~Petronzio and E.~Zavattini,
  Phys.\ Lett.\  B {\bf 175} (1986) 359.

\bibitem{Raffelt:1987im}
  G.~Raffelt and L.~Stodolsky,
  Phys.\ Rev.\  D {\bf 37} (1988) 1237.

\bibitem{Gies:2006ca}
  H.~Gies, J.~Jaeckel and A.~Ringwald,
  Phys.\ Rev.\ Lett.\  {\bf 97} (2006) 140402
  [arXiv:hep-ph/0607118].

{
\bibitem{Heinzl:2006xc}
  T.~Heinzl, B.~Liesfeld, K.~U.~Amthor, H.~Schwoerer, R.~Sauerbrey and A.~Wipf,
  Opt.\ Commun.\  {\bf 267}, 318 (2006)
  [arXiv:hep-ph/0601076].

\bibitem{Ruoso:1992nx}
  G.~Ruoso {\it et al.} [BFRT Collaboration],
  Z.\ Phys.\  C {\bf 56} (1992) 505.



\bibitem{Okun:1982xi}
L.~B. Okun,
\newblock Sov. Phys. JETP {\bf 56}, 502 (1982).

\bibitem{Sikivie:1983ip}
  P.~Sikivie,
  Phys.\ Rev.\ Lett.\  {\bf 51} (1983) 1415
  [Erratum-ibid.\  {\bf 52} (1984) 695].

\bibitem{Anselm:1986gz}
  A.~A.~Anselm,
  Yad.\ Fiz.\  {\bf 42} (1985) 1480.


\bibitem{Gasperini:1987da}
  M.~Gasperini,
  Phys.\ Rev.\ Lett.\  {\bf 59} (1987) 396.

\bibitem{VanBibber:1987rq}
K.~Van~Bibber, N.~R. Dagdeviren, S.~E. Koonin, A.~Kerman, and H.~N. Nelson,
\newblock Phys. Rev. Lett. {\bf 59}, 759 (1987).


\bibitem{Dupays:2005xs}
  A.~Dupays, C.~Rizzo, M.~Roncadelli and G.~F.~Bignami,
  Phys.\ Rev.\ Lett.\  {\bf 95}, 211302 (2005)
  [arXiv:astro-ph/0510324].
}

\bibitem{Mirizzi:2007hr}
  A.~Mirizzi, G.~G.~Raffelt and P.~D.~Serpico,
  arXiv:0704.3044 [astro-ph].

\bibitem{Fairbairn:2007vj}
  M.~Fairbairn, S.~N.~Gninenko, N.~V.~Krasnikov, V.~A.~Matveev, T.~I.~Rashba, A.~Rubbia and S.~Troitsky,
  arXiv:0706.0108 [hep-ph].
 

\bibitem{Ehret:2007cm}
  K.~Ehret {\it et al.} [ALPS collaboration],
  arXiv:hep-ex/0702023.


\bibitem{Rizzo:Patras}
C.~Rizzo for the [BMV Collaboration],
2nd ILIAS-CERN-CAST Axion Academic Training 2006,
http://cast.mppmu.mpg.de/

\bibitem{GammeV}
GammeV Collaboration, http://gammev.fnal.gov/

\bibitem{Baker:Patras}
K.~Baker for the [LIPSS Collaboration],
2nd ILIAS-CERN-CAST Axion Academic Training 2006,
http://cast.mppmu.mpg.de/

\bibitem{OSQAR}
P. Pugnat {\em et al.} [OSQAR Collaboration],
CERN-SPSC-2006-035, CERN-SPSC-P-331.

\bibitem{Cantatore:Patras}
G.~Cantatore for the [PVLAS Collaboration],
2nd ILIAS-CERN-CAST Axion Academic Training 2006,
http://cast.mppmu.mpg.de/


\bibitem{Ringwald:2006rf}
  A.~Ringwald,
  arXiv:hep-ph/0612127.

\bibitem{Battesti:2007um}
  R.~Battesti {\it et al.},
  arXiv:0705.0615 [hep-ex].

{
\bibitem{Baier}
R.~Baier and P.~Breitenlohner, 
Act.~Phys.~Austriaca {\bf 25}, 212 (1967); 
Nuov.~Cim.~B {\bf 47} 117 (1967).
}

\bibitem{Adler:1971wn}
  S.~L.~Adler,
  Annals Phys.\  {\bf 67} (1971) 599.

\bibitem{Adler:2006zs}
  S.~L.~Adler,
  J.\ Phys.\ A  {\bf 40} (2007) F143
  [arXiv:hep-ph/0611267].

\bibitem{Biswas:2006cr}
  S.~Biswas and K.~Melnikov,
  Phys.\ Rev.\  D {\bf 75} (2007) 053003
  [arXiv:hep-ph/0611345].



\bibitem{Ahlers:2006iz}
M.~Ahlers, H.~Gies, J.~Jaeckel, and A.~Ringwald,
\newblock Phys. Rev. {\bf D75}, 035011 (2007), hep-ph/0612098.


\bibitem{Raffelt:1996}
G.~G.~Raffelt,
Stars As Laboratories For Fundamental Physics:
The Astrophysics of Neutrinos, Axions, and other Weakly Interacting Particles,
University of Chicago Press, Chicago, 1996.



\bibitem{Davidson:2000hf}
  S.~Davidson, S.~Hannestad and G.~Raffelt,
  JHEP {\bf 0005} (2000) 003
  [arXiv:hep-ph/0001179].



\bibitem{Masso:2005ym}
  E.~Masso and J.~Redondo,
  JCAP {\bf 0509} (2005) 015
  [arXiv:hep-ph/0504202].

\bibitem{Jain:2005nh}
  P.~Jain and S.~Mandal,
  Int.\ J.\ Mod.\ Phys.\  D {\bf 15} (2006) 2095
  [arXiv:astro-ph/0512155].

\bibitem{Masso:2006gc}
  E.~Masso and J.~Redondo,
  Phys.\ Rev.\ Lett.\  {\bf 97} (2006) 151802
  [arXiv:hep-ph/0606163].

\bibitem{Abel:2006qt}
  S.~A.~Abel, J.~Jaeckel, V.~V.~Khoze and A.~Ringwald,
  arXiv:hep-ph/0608248.


\bibitem{Jaeckel:2006xm}
  J.~Jaeckel, E.~Masso, J.~Redondo, A.~Ringwald and F.~Takahashi,
  Phys.\ Rev.\  D {\bf 75} (2007) 013004
  [arXiv:hep-ph/0610203].



\bibitem{Mohapatra:2006pv}
  R.~N.~Mohapatra and S.~Nasri,
  Phys.\ Rev.\ Lett.\  {\bf 98} (2007) 050402
  [arXiv:hep-ph/0610068].

\bibitem{Jain:2006ki}
  P.~Jain and S.~Stokes,
  arXiv:hep-ph/0611006.

\bibitem{Foot:2007cq}
  R.~Foot and A.~Kobakhidze,
  Phys.\ Lett.\  B {\bf 650} (2007) 46
  [arXiv:hep-ph/0702125].

\bibitem{Brax:2007ak}
  P.~Brax, C.~van de Bruck and A.~C.~Davis,
  arXiv:hep-ph/0703243.

\bibitem{Kim:2007wj}
  J.~E.~Kim,
  arXiv:0704.3310 [hep-ph].


\bibitem{Melchiorri:2007sq}
  A.~Melchiorri, A.~Polosa and A.~Strumia,
  arXiv:hep-ph/0703144.


\bibitem{Holdom:1985ag}
  B.~Holdom,
  Phys.\ Lett.\  B {\bf 166} (1986) 196.

\bibitem{Dienes:1996zr}
  K.~R.~Dienes, C.~F.~Kolda and J.~March-Russell,
  Nucl.\ Phys.\  B {\bf 492} (1997) 104
  [arXiv:hep-ph/9610479].

\bibitem{Lust:2003ky}
  D.~Lust and S.~Stieberger,
  arXiv:hep-th/0302221.

\bibitem{Abel:2003ue}
  S.~A.~Abel and B.~W.~Schofield,
  Nucl.\ Phys.\  B {\bf 685}, 150 (2004)
  [hep-th/0311051].

\bibitem{Abel:2004rp}
  S.~Abel and J.~Santiago,
  J.\ Phys.\ G {\bf 30}, R83 (2004)
  [hep-ph/0404237].



\bibitem{Batell:2005wa}
  B.~Batell and T.~Gherghetta,
  Phys.\ Rev.\  D {\bf 73} (2006) 045016
  [arXiv:hep-ph/0512356].

\bibitem{Blumenhagen:2006ux}
  R.~Blumenhagen, S.~Moster and T.~Weigand,
  Nucl.\ Phys.\  B {\bf 751} (2006) 186
  [arXiv:hep-th/0603015].


\bibitem{Davidson:1991si}
S.~Davidson, B.~Campbell, and D.~C. Bailey,
\newblock Phys. Rev. {\bf D43}, 2314 (1991).

\bibitem{Williams:1971ms}
  E.~R.~Williams, J.~E.~Faller and H.~A.~Hill,
  Phys.\ Rev.\ Lett.\  {\bf 26} (1971) 721.

\bibitem{Bartlett:1988yy}
  D.~F.~Bartlett and S.~Loegl,
  Phys.\ Rev.\ Lett.\  {\bf 61} (1988) 2285.

\bibitem{Popov:1999}
  V.~Popov, 
  Turk.\ J.\ Phys.\  {\bf 23} (1999) 943.

\bibitem{Wilczek:1977pj}
F.~Wilczek,
\newblock Phys. Rev. Lett. {\bf 40}, 279 (1978).

\bibitem{Weinberg:1977ma}
S.~Weinberg,
\newblock Phys. Rev. Lett. {\bf 40}, 223 (1978).


\bibitem{Schwinger:1951nm}
J.~S. Schwinger,
\newblock Phys. Rev. {\bf 82}, 664 (1951).

\bibitem{Erber:1966vv}
T.~Erber,
\newblock Rev. Mod. Phys. {\bf 38}, 626 (1966).

\bibitem{Dittrich:2000zu}
W.~Dittrich and H.~Gies,
%
Springer Tracts Mod.\ Phys.\  {\bf 166}, 1 (2000).

\bibitem{Shore:2007um}
  G.~M.~Shore,
  arXiv:hep-th/0701185.


\bibitem{Liao:2007nu}
Y.~Liao,
\newblock (2007), arXiv:0704.1961 [hep-ph].


\bibitem{DiPiazza:2006pr}
  A.~Di Piazza, K.~Z.~Hatsagortsyan and C.~H.~Keitel,
  Phys.\ Rev.\ Lett.\  {\bf 97}, 083603 (2006)
  [arXiv:hep-ph/0602039].

\bibitem{Marklund:2006my}
  M.~Marklund and P.~K.~Shukla,
  Rev.\ Mod.\ Phys.\  {\bf 78}, 591 (2006)
  [arXiv:hep-ph/0602123].

\bibitem{Gies:2006hv}
  H.~Gies, J.~Jaeckel and A.~Ringwald,
  Europhys.\ Lett.\  {\bf 76} (2006) 794
  [arXiv:hep-ph/0608238].



\bibitem{Tsai:1975iz}
W.-y. Tsai and T.~Erber,
\newblock Phys. Rev. {\bf D12}, 1132 (1975).

\bibitem{Tsai:1974fa}
W.-y. Tsai and T.~Erber,
\newblock Phys. Rev. {\bf D10}, 492 (1974).

\bibitem{Toll:1952rq}
J.~S. Toll,
\newblock {\em The Dispersion relation for light and its application to
  problems involving electron pairs},
\newblock PhD thesis,
\newblock RX-1535.

\bibitem{Klepikov:1954}
N.~P. Klepikov,
\newblock Zh. Eksp. Teor. Fiz. {\bf 26} (1954).

\bibitem{Baier:1967}
V.~Baier and V.~Katkov,
\newblock Zh. Eksp. Teor. Fiz. {\bf 53}, 1478 (1967).

\bibitem{Klein:1968}
J.~J. Klein,
\newblock Rev. Mod. Phys. {\bf 40}, 523 (1968).


\bibitem{Daugherty:1984tr}
J.~K.~Daugherty and A.~K.~Harding,
Astrophys.\ J.\  {\bf 273}, 761 (1983).

\end{thebibliography}
\end{document}